%% file: main.tex
\documentclass[sigconf,screen,nonacm]{acmart}

\usepackage{enumitem}
\usepackage{adjustbox}
\usepackage{xspace}
\usepackage{pifont}
\usepackage{amsmath}
\usepackage{amssymb}
\usepackage{tcolorbox}
\usepackage{multirow}
\usepackage{wrapfig}
\usepackage{graphicx}
\usepackage{svg}
\usepackage[table,xcdraw]{xcolor}

\usepackage[ruled,vlined,linesnumbered]{algorithm2e}

\usepackage{subcaption}

\newcommand{\kai}[1]{\textcolor{black}{#1}}

\newcommand{\chen}[1]{\textcolor{red}{#1}}

\definecolor{myLightBlue}{RGB}{242,242,253}   
\definecolor{myBlueLine}{RGB}{128,127,255}    

\newtcolorbox{answerbox_round}{
  colback=myLightBlue,
  colframe=myBlueLine,
  boxrule=0pt,
  leftrule=3pt,
  arc=2pt,
  left=0pt,
  right=0pt,
  top=1pt,
  bottom=1pt,
  before skip=5pt,           
  after skip=5pt,            
}

\newtcolorbox{answerbox}{
  colback=myLightBlue,        
  colframe=myBlueLine,        
  boxrule=0pt,                
  leftrule=3pt,               
  arc=1pt,                    
  left=3pt, right=3pt,        
  top=3pt, bottom=3pt,        
  sharp corners,              
  width=\linewidth,           
  before skip=7pt,           
  after skip=6pt,            
}

\usepackage{xcolor}

\definecolor{box0color}{RGB}{255, 0, 0}    
\definecolor{box1color}{RGB}{230, 74, 25}    
\definecolor{box2color}{RGB}{255, 193, 7}    
\definecolor{box3color}{RGB}{13, 110, 253}   
\definecolor{box4color}{RGB}{111, 66, 193}   
\definecolor{box7color}{RGB}{25, 135, 84}    
\definecolor{box8color}{RGB}{255, 193, 7}    

\definecolor{myLightBlue}{RGB}{242,242,253}    
\definecolor{myLightGreen}{RGB}{242,251,242}   
\definecolor{myLightYellow}{RGB}{255,250,235}  
\definecolor{myLightOrange}{RGB}{255,244,235}  
\definecolor{myLightPurple}{RGB}{248,242,255}  
\definecolor{myLightGray}{RGB}{245,246,248}    
\definecolor{myLightPink}{RGB}{255,242,248}    
\definecolor{myLightBlueGray}{RGB}{244,244,251}

\AtBeginDocument{%
  }

\setcopyright{acmlicensed}
\copyrightyear{2018}
\acmYear{2018}
\acmDOI{XXXXXXX.XXXXXXX}
\acmConference[Conference acronym 'XX]{Make sure to enter the correct
  conference title from your rights confirmation email}{June 03--05,
  2018}{Woodstock, NY}
\acmISBN{978-1-4503-XXXX-X/2018/06}

\begin{document}

\title{A11YRepair: Bridging Web Accessibility Barriers via Knowledge-Enhanced Divide-and-Conquer Repair}

\author{Kai Huang}
\email{kai-kevin.huang@tum.de}
\affiliation{%
  \institution{Technical University of Munich}
  \country{Germany}
}

\author{Ling Zhu}
\email{julialing.zhu@tum.de}
\affiliation{%
  \institution{Technical University of Munich}
  \country{Germany}
}

\author{Jian Zhang}
\email{jian\_zhang@ntu.edu.sg}
\affiliation{%
  \institution{Nanyang Technological University}
  \country{Singapore}
}

\author{Xiaofei Xie}
\email{xfxie@smu.edu.sg}
\affiliation{%
  \institution{Singapore Management University}
  \country{Singapore}
}

\author{Chunyang Chen}
\email{chun-yang.chen@tum.de}
\affiliation{%
  \institution{Technical University of Munich}
  \country{Germany}
}

\renewcommand{\shortauthors}{Huang et al.}

\begin{abstract}

\kai{Web accessibility (A11Y), which ensures web content is perceivable and usable for users with disabilities, is a critical requirement for modern web applications. Yet existing tooling overwhelmingly focuses on detecting A11Y violations rather than repairing them.}
Automated program repair (APR) techniques appear promising for this setting, but our study shows that state-of-the-art APR systems perform poorly when applied to real-world A11Y violations. Unlike conventional sparse-bug scenarios, web A11Y issues often manifest as multiple structurally related violations per page, requiring coordinated edits across multiple files. 
Existing repair systems fail to manage this multi-fault scale, as they handle each bug individually without considering their relationships or incorporating domain rules such as the Web Content Accessibility Guidelines (WCAG).

We propose A11YRepair,
an LLM-based framework 
designed 
specifically 
for web A11Y repair. A11YRepair introduces a divide-and-conquer workflow that first clusters violations requiring coordinated edits to reduce redundant localization, and then decomposes each cluster by root cause so the LLM can generate focused and consistent patches. The framework further incorporates WCAG-driven knowledge to strengthen domain awareness during both fault localization and patch synthesis. 

To support systematic evaluation, we construct A11YBench, a benchmark of 60 real-world web projects collected from GitHub. Experimental results show that A11YRepair achieves higher repair effectiveness and lower cost than state-of-the-art baselines, and ablation studies confirm the importance of its divide-and-conquer design and selective domain knowledge integration. 
\kai{Specifically, patches generated by A11YRepair have been merged into open-source projects from Google, Microsoft, Facebook, IBM, 
K8s, Docker, 
and Alibaba, demonstrating its practical value
in real-world
dev scenarios.
}
\end{abstract}

%
%





\maketitle

\input{section/1_Introduction}
\input{section/2_3_Study}
\input{section/4_Approach}
\input{section/5_Setup}

\input{section/6_Evaluation}

\input{section/7_Threats}

\input{section/8_Related}
\input{section/9_Conclusion}
\input{section/10_Data}

\bibliographystyle{ACM-Reference-Format}
\bibliography{sample-base}

\end{document}

%% file: section/1_Introduction.tex
\section{Introduction}


\begin{figure}[t]
    \centering
    \includegraphics[width=1.0\linewidth, trim=0 8 0 0, clip]
    {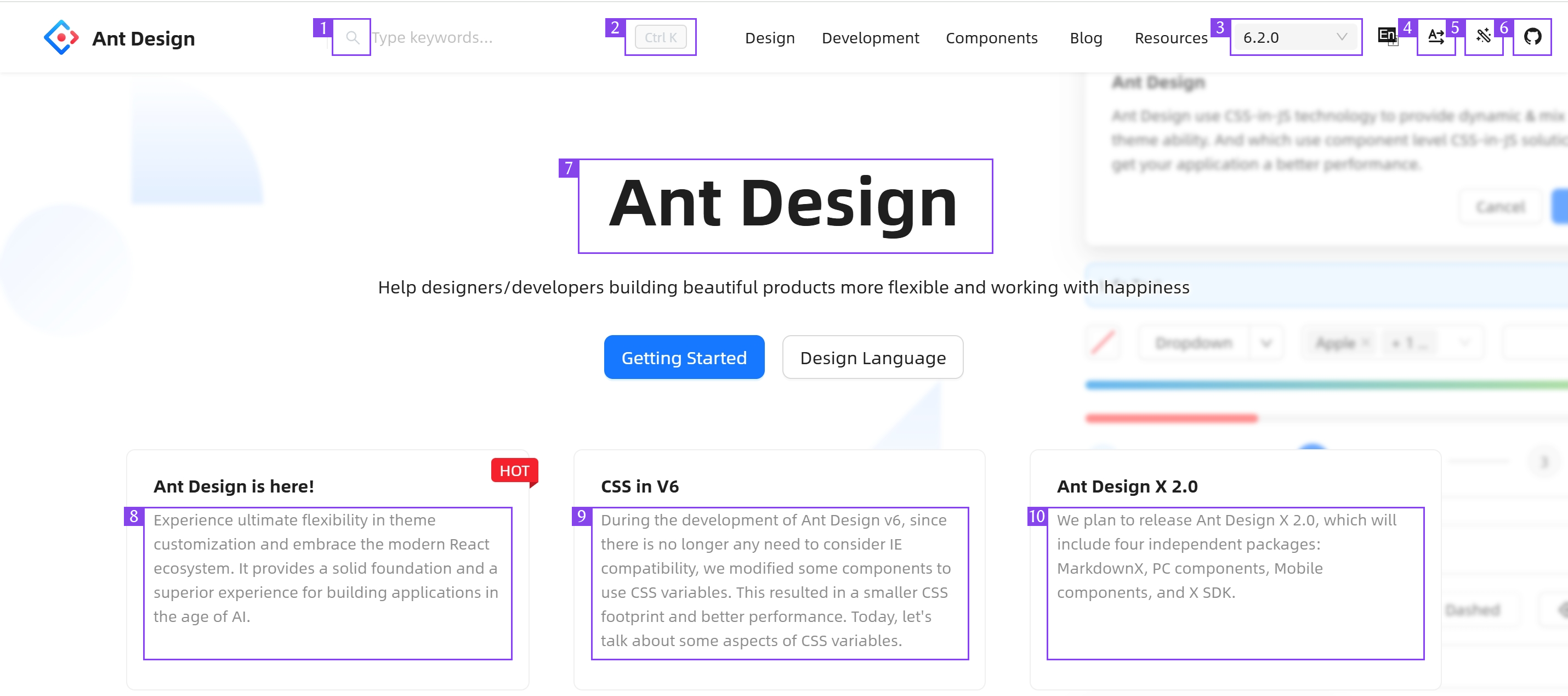}
    \vspace{-16pt}
    \caption{Multiple accessibility violations detected on the \href{https://ant.design/}{Ant Design homepage} by using the \href{https://www.ibm.com/able/toolkit/tools/}{IBM Accessibility Checker}.
    \small{
    The numbered boxes highlight different violations: 
    \textcolor{box4color}{\textbf{\ding{182}}} Missing accessible name for SVG element; \textcolor{box4color}{\textbf{\ding{183}}} Insufficient contrast (1.42:1 for 12px text); \textcolor{box4color}{\textbf{\ding{184}}} Unlabeled combobox with invalid ARIA role; \textcolor{box4color}{\textbf{\ding{185}\ding{186}\ding{187}}} Button elements with ignored descendant img roles; \textcolor{box4color}{\textbf{\ding{188}}} Content outside landmark regions; \textcolor{box4color}{\textbf{\ding{189}\ding{190}\ding{191}}} Insufficient contrast (3.35:1 for 14px text). 
    }
    }
    \Description{The overview of A11YRepair.}
    \label{fig:web_a11y_example}
    \vspace{-12pt}
\end{figure}

The Web has become an essential part of everyday life, yet accessibility (A11Y) remains one of its most persistent challenges. 
Despite decades of progress and widely adopted standards such as the Web Content Accessibility Guidelines (WCAG)~\cite{WCAG22}, recent audits still find that more than 94\% of high-traffic websites contain accessibility violations~\cite{WebAIMMillion}. 
These issues not only hinder millions of users with disabilities from fully participating online but also impose growing compliance and maintenance costs on developers and organizations~\cite{A11yDev,Ma11y,GenA11y,Web_Ad_A11Y}. 
As illustrated in Figure~\ref{fig:web_a11y_example}, even Ant Design~\cite{Ant_Design_Github}, a popular UI framework on GitHub with 97k+ stars and maintained by Ant Group's development team, exhibits numerous accessibility violations on its official homepage~\cite{Ant_Design_Web}. 
\kai{
These violations span multiple WCAG criteria and create real barriers for users with disabilities: low contrast text makes content unreadable for people with moderately low vision, missing landmarks prevent efficient navigation and skipping of repeated content, unlabeled img elements remain invisible to screen readers, and improper ARIA roles can cause duplicate announcements that confuse assistive technology users. 
}
This example underscores a troubling reality: accessibility is often overlooked even in high-profile, professionally maintained projects, and multiple diverse violations commonly co-occur on a single page.
Existing accessibility checkers, such as IBM’s Accessibility Checker and Deque’s axe-core~\cite{IBM_A11YChecker,Axe_Core}, have improved detection, but they stop short of automated repair.
Developers must still interpret WCAG rules manually, making remediation time-consuming, inconsistent, and error-prone, especially as modern web pages may include dozens of interdependent elements with accessibility flaws. 
%

Recent research has called for a shift from detection to intervention~\cite{LLM_a11y_study}, arguing that identifying violations alone is insufficient to achieve real accessibility. While large language models (LLMs) have advanced automated program repair (APR)~\cite{SWE_agent,AutoCodeRover,OpenHands}, their ability to repair web accessibility violations remains largely unexplored. 
This gap arises because accessibility issues differ fundamentally from traditional bug scenarios:
they are highly clustered~\cite{WebAIMMillion}, visually grounded~\cite{TaskAudit}, and governed by domain-specific standards~\cite{GenA11y}.
For example, WCAG~\cite{WCAG22} defines 86 \textit{Success Criteria} (\textit{Requirements}) and 595 \textit{Techniques}~\cite{WCAG22_Techniques} for ensuring web accessibility.

To investigate this, we conduct a preliminary study of repair agents for web A11Y repair. Specifically, we design and compare three repair strategies that vary in how they handle violation reports (e.g., jointly vs. individually) and incorporate domain knowledge.
Our analysis reveals three key challenges. 
\textcolor{box4color}{\textbf{\ding{182}}} When repair agents process entire violation reports at once, they struggle to plan and coordinate across multiple violations, leading to incomplete or incoherent patches. 
\textcolor{box4color}{\textbf{\ding{183}}} When handling each violation separately, they achieve higher precision but suffer from excessive repetition, side effects, and inflated cost. 
\textcolor{box4color}{\textbf{\ding{184}}} Although WCAG knowledge is crucial for complex violations, indiscriminate retrieval often wastes context and introduces noise.
These observations motivate the need for a repair framework that balances task granularity, captures violation relationships, and selectively integrates domain knowledge.

Building on these observations, we identify several key design requirements for web accessibility repair. 
\textcolor{box4color}{\textbf{\ding{182}}} Repair frameworks must adapt task granularity to the multi-fault nature of web pages, enabling structured planning while avoiding both holistic overload and redundant per-violation repairs.
\textcolor{box4color}{\textbf{\ding{183}}} Repair systems need to model cross-violation and component-level relationships to prevent duplicated effort and unintended side effects.
\textcolor{box4color}{\textbf{\ding{184}}} Domain knowledge such as WCAG guidelines should be integrated selectively, providing precise guidance for complex, standards-sensitive violations without unnecessarily increasing context overhead for simple cases.
Together, these requirements motivate a new repair paradigm that aligns LLM-based automation with the practical and structural characteristics of real-world web accessibility engineering.

Guided by these insights, we propose \textbf{A11YRepair}, an LLM-based framework for automated repair of web accessibility violations.
A11YRepair adopts a goal-oriented divide-and-conquer strategy that explicitly adapts task granularity by grouping multiple related violations, rather than repairing each violation in isolation.
Violations are grouped based on component structure, code locality, and WCAG criteria, enabling the framework to synthesize coordinated repairs for multiple related violations within the same code region.
This design prevents redundant localization and mitigates side effects that commonly arise from individually 
generated patches.
To support accurate fault identification, A11YRepair’s \textit{fault localization} module integrates DOM-structure retrieval with visual context analysis to pinpoint fault-relevant files and components.
Its \textit{patch generation} loop then produces minimal, root-cause–aware edits that are jointly applied to grouped violations 
and iteratively validated through re-checking.
In addition, A11YRepair employs a \textit{selective WCAG-driven reasoning mechanism} that retrieves guideline content only when violations require quantitative thresholds or semantic interpretation, avoiding unnecessary context overhead for simple cases.
Together, these designs enable A11YRepair to reason globally while acting locally, achieving an effective balance between repair precision, cost efficiency, and standards compliance.


We evaluate A11YRepair on \textbf{A11YBench},
a benchmark of 60 real-world GitHub web projects with 8,886 violations.
A11YRepair resolves 76.8\% of violations at an average cost of \$0.018 per issue, outperforming strong baselines such as GUIRepair in both effectiveness and cost.
Notably, patches generated by A11YRepair have been merged by open-source projects from Google, Microsoft, Facebook, IBM, etc.
In summary, we make the following contributions:


\begin{itemize}[leftmargin=0.4cm]
    \item We built A11YBench, the first repo-level benchmark for web accessibility repair, comprising 60 real-world GitHub web projects.
    \item We propose A11YRepair, the first LLM-based source-level web accessibility violation repair framework that combines divide-and-conquer grouping
    and selective WCAG-guided reasoning. 
    \item We show that A11YRepair achieves SOTA performance and practical value,
    and we release artifacts to support future research~\cite{A11YRepair_link}. 
\end{itemize}

%% file: section/2_3_Study.tex
\section{Benchmark Construction}
\label{sec:bench+study}

We construct A11YBench by collecting real-world GitHub repositories that implement production web applications using \textit{JavaScript} or \textit{TypeScript}. Detailed construction procedures and collected projects are documented on~\cite{A11YBench_link}.
\kai{Overall, A11YBench consists of 60 web projects, encompassing 147 web pages and 8,886 accessibility violations spanning 45 distinct violation types, as detected by the IBM Accessibility Checker.}
The projects vary substantially in size, from 123 to 43,198 source files and from 3,610 to 1,555,532 lines of code, covering both lightweight documentation sites and large production-grade applications. 
This scale ensures that A11YBench reflects the structural and technical diversity of modern web ecosystems. 
Similar to SWE-bench~\cite{SWEbench}, we divide A11YBench into two sets: \textbf{Lite} and \textbf{Full}, to balance evaluation cost.
A11YBench-Lite includes 10 randomly selected repositories for quick testing, A11YBench-Full comprises all 60 repositories for comprehensive evaluation.
This design enables flexible benchmarking across lightweight and large-scale settings~\cite{SWEbench}.

\section{Motivating Study}

To understand the limitations of existing repair tools in real-world web scenarios, we conduct a lightweight motivating study to identify key challenges that motivate the design of A11YRepair.

\subsection{Study Setup}
\label{sec:exp-setup}

\subsubsection{Repair Workflow}
We design three input configurations
to investigate existing repair systems in the web accessibility scenario:

\begin{answerbox_round}
\footnotesize{
\begin{itemize}[leftmargin=0.4cm]
    \item \textit{\textbf{Basic Repair Strategy}.}  
    The system receives an entire A11Y report at once, containing multiple violations.  
    This strategy evaluates the agent’s ability to plan autonomously and coordinate multiple violations within a single context.
    \item \textit{\textbf{Iterative Repair Strategy}.}  
    The system processes each violation instance individually, allowing focused reasoning on individual issues by decomposing the overall complex problem space into simpler, isolated repair sub-tasks directly.
    \item \textit{\textbf{WCAG Guided Repair Strategy}.}  
    This variant extends the \textit{Iterative Repair Strategy} by incorporating relevant WCAG guidelines for each violation instance, enabling us to assess the impact of domain knowledge on repair effectiveness.
\end{itemize}
}
\end{answerbox_round}

\subsubsection{Systems and Metrics}
Given that existing web A11Y repair tools~\cite{Access,AccessGuru,CodeA11Y} cannot apply fixes at the repo-level source code, we select 3 general LLM-based repair systems: \textit{SWE-agent}~\cite{SWE_agent}, \textit{OpenHands}~\cite{OpenhandsVersa}, and 
\textit{GUIRepair}~\cite{GUIRepair}.
They represent the top-performing repair systems in the multimodal issue repair~\cite{Leaderboard}.
In the evaluation metrics, we report \textit{Violation Solve Rate} ($R_{\text{solve}}$), \textit{Side Effect Count} ($N_{\text{side}}$), and \textit{Token Usage Cost} ($C_{\text{total}}$), 
as defined in Section~\ref{sec:EvaluationMetrics}. 

\subsubsection{Task Instances Selection}
We randomly select 5 repos from A11YBench: 
\textit{\href{https://github.com/TheAlgorithms/website}{algorithms}}, \textit{\href{https://github.com/carbon-design-system/carbon-website}{carbon-design}}, \textit{\href{https://github.com/tailwindlabs/tailwindcss.com}{tailwindcss}}, 
\textit{\href{https://github.com/lynx-family/lynx-website}{lynx}}, \textit{\href{https://github.com/electron/website}{electron}}. To balance the cost, 20 webpages are retained randomly, containing 250 violations spanning 22 distinct WCAG~2.2 \textit{Success Criteria}. 

\begin{figure*}[t]
    \centering
    \vspace{-2pt}
    \includegraphics[width=1.0\linewidth, trim=336 152 336 152, clip]{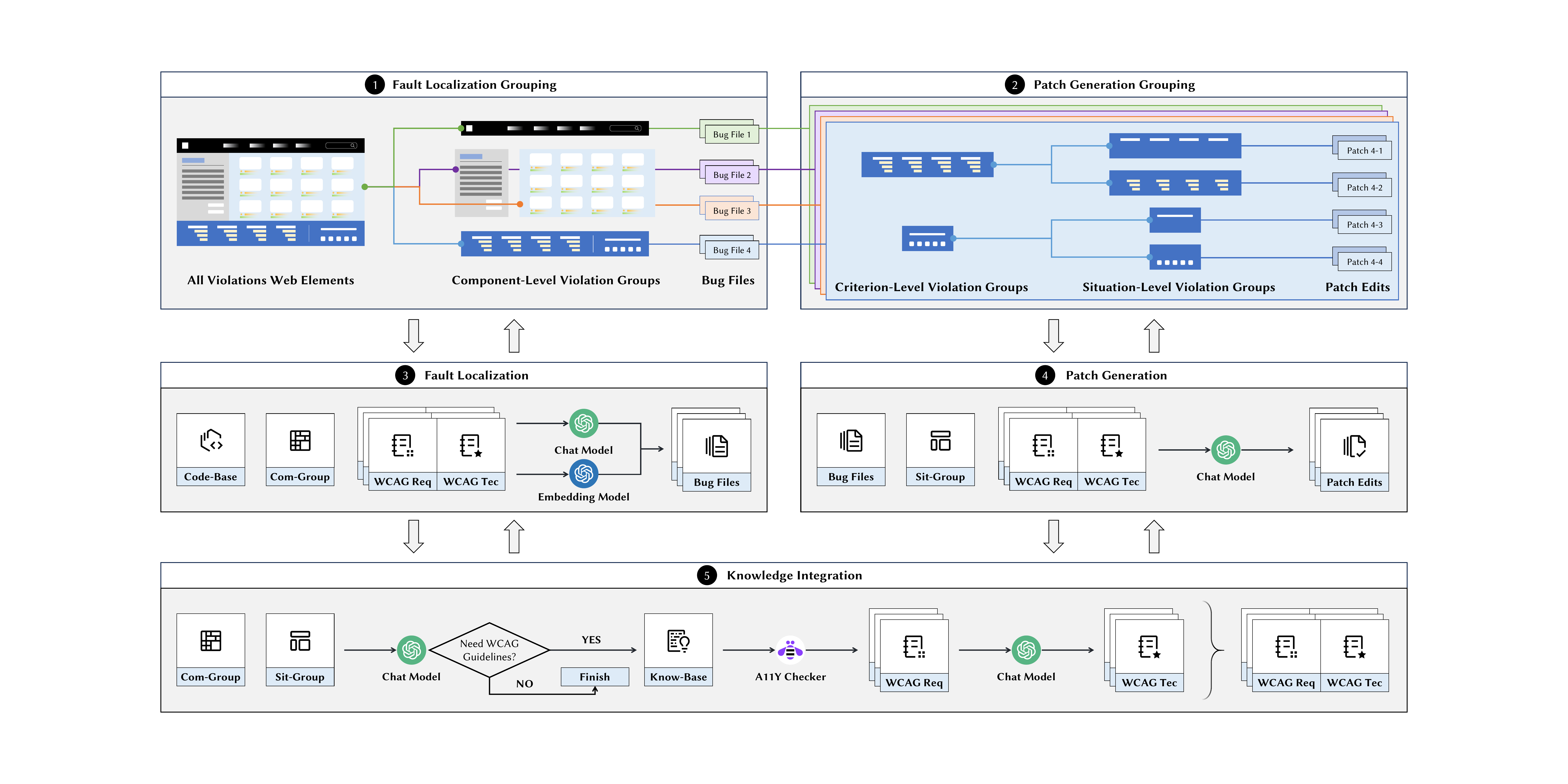}
    \vspace{-13pt}
    \caption{The overview of A11YRepair. 
    \small
    {\ding{182}-\ding{183} It first groups all violations at the component and situation levels for fault localization and patch generation. \ding{184}-\ding{185} Then, it leverages chat and embedding models to locate buggy files and synthesize patch edits. \ding{186} Meanwhile, knowledge integration module selectively incorporates WCAG Requirements~\cite{IBM_A11Y_Requirements} and Techniques~\cite{WCAG22_Techniques} for fault localization and patch generation.} 
    }
    \Description{The overview of A11YRepair.}
    \label{fig:a11yrepair_workflow}
    \vspace{-10pt}
\end{figure*}

\subsection{Results and Analysis}

\subsubsection{Basic Repair Strategy}
Under the Basic Repair Strategy, repair system processes the entire issue report with all violations in a single interaction. 
As shown in Table~\ref{tab:basic_and_iter_repair_strategy}, although repair systems incur low costs, their repair capability remains constrained. Even the best-performing GUIRepair, reduces violations by only 31.52\%. 

This limitation is attributable to the inherently multi-fault nature of web accessibility repair. As illustrated in Figure~\ref{fig:web_a11y_example}, accessibility violations on a single webpage are often numerous and widely distributed~\cite{WebAIMMillion}. In A11YBench, 85.03\% (125/147) of webpages contain more than 10 violations, and 12.24\% (18/147) contain over 100 violations, indicating that large-scale multi-fault scenarios are common rather than exceptional.
Such problem scale poses simultaneous challenges to both fault localization and patch generation. 
When a complete violation report is provided as input, repair systems are required to identify multiple faulty locations across different files and apply diverse fixes within a single generation. However, existing repair agents lack explicit mechanisms for task decomposition and hierarchical planning. As a result, they often fail to localize all relevant bug files and produce incoherent or incomplete patches when multiple fixes must be synthesized at once.


These observations suggest that repair systems need explicit mechanisms for hierarchical planning and simplified patch synthesis to manage large and complex repair tasks effectively.

\begin{answerbox}
\small{
\textit{\textbf{Finding 1:} The Basic Repair Strategy achieves low repair effectiveness despite minimal computational cost. Current LLM-based repair agents lack robust task planning and hierarchical reasoning, resulting in inaccurate fault localization and incoherent patches under multi-fault conditions.} 
}
\end{answerbox}

\input{table/basic_and_iterative_result_small}

\subsubsection{Iterative Repair Strategy}

Unlike the basic strategy, which receives all violations at once, 
iterative strategy processes each violation individually.
As shown in Table~\ref{tab:basic_and_iter_repair_strategy}, it achieves a higher resolve rate than basic strategy, but at the cost of reduced efficiency and increased side effects.
We illustrate these limitations with examples.


\textbf{Repairing centrally repairable violations individually.}
\setlength{\columnsep}{10pt}
\begin{wrapfigure}{l}{0.574\columnwidth} 
\vspace{-1pt}
    \begin{center}
        \subfloat[Fix before: 3 detected Vio. \textcolor{box4color}{\textbf{\ding{182}}} \textcolor{box4color}{\textbf{\ding{183}}} \textcolor{box4color}{\textbf{\ding{184}}}. 
        \label{fig:example_footer}]{\includegraphics[width=\linewidth, trim=225 200 225 190, clip]{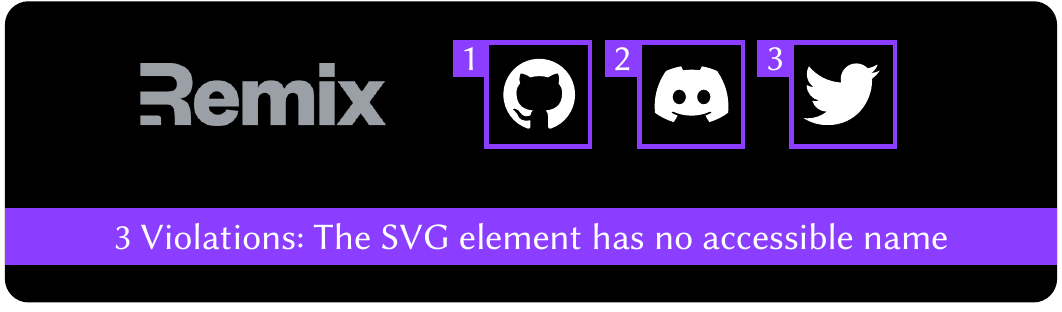}}
    \end{center}
    \begin{center}
        \subfloat[Patch of 3 same repair pattern.
        \label{fig:example_footer_code}]{\includegraphics[width=\linewidth, trim=258 97 258 36, clip]{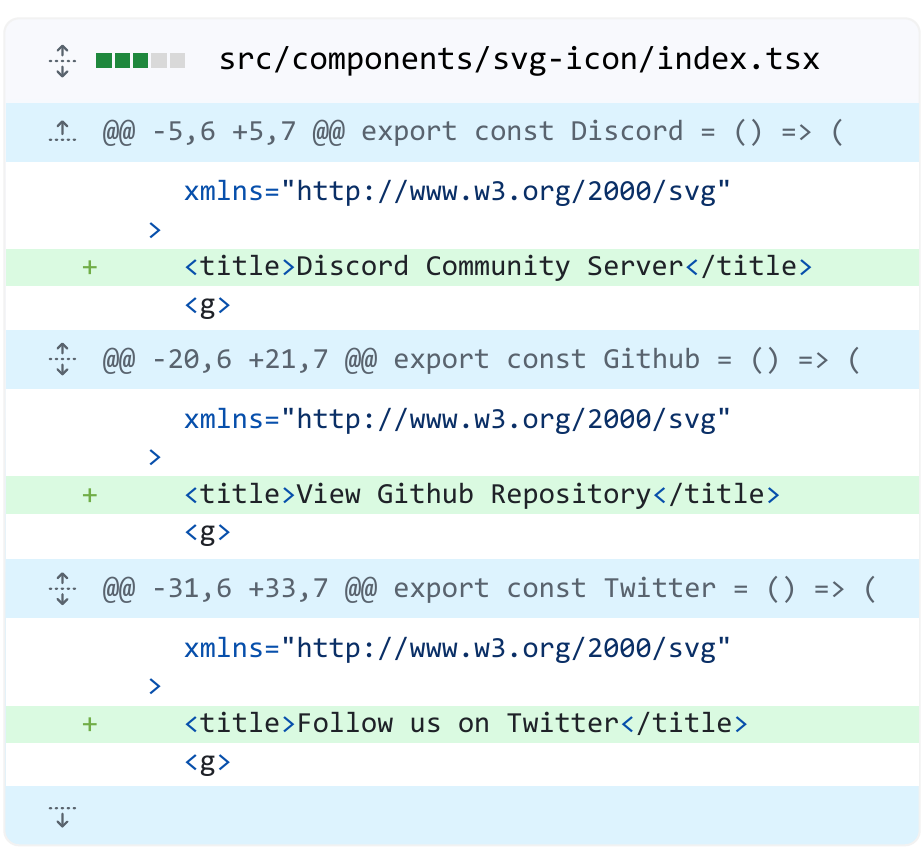}}
    \end{center}
    \vspace{-8pt}
    \caption{A motivating example of centrally repairable violations. 
    } 
    \label{fig:knowledge_mining_example}
\vspace{-4pt}
\end{wrapfigure}
\noindent
Figure~\ref{fig:example_footer} shows three SVG elements that lack textual alternatives, preventing screen reader users from understanding their navigation purposes.
As shown in Figure~\ref{fig:example_footer_code}, the developer can resolve all three violations by applying the same repair behavior—adding descriptive text to a shared component file in a single repair action.
These violations share the same root cause and repair pattern, 
meaning they can be repaired centrally.
However, under the iterative repair strategy, the LLM repairs each violation individually, repeatedly localizing the same component and synthesizing identical patches, which leads to redundant queries and unnecessary overhead in practice.

\textbf{Side effects from ignoring inter-component dependencies.}

\begin{wrapfigure}{l}{0.574\columnwidth} 
    \vspace{-12pt}
        \begin{center}
            \subfloat[Fix after: 2 new introduced Vio. \textcolor{box1color}{\textbf{\ding{182}}} \textcolor{box1color}{\textbf{\ding{183}}}. 
            \label{fig:example_footer_after}]{\includegraphics[width=1.0\linewidth, trim=225 200 225 190, clip]{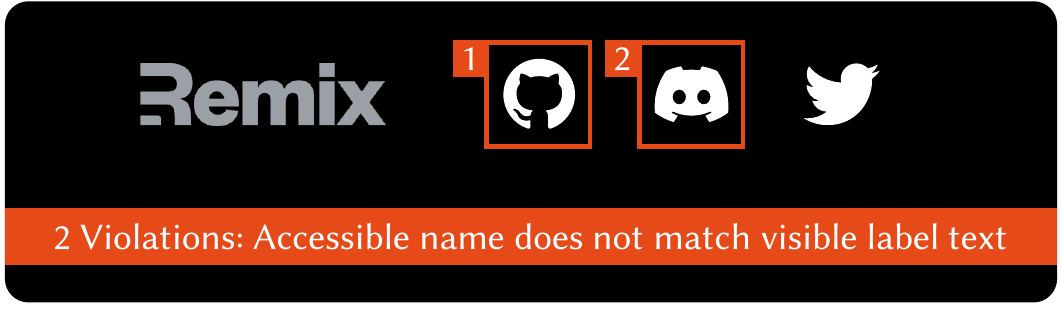}}
            \end{center}
    \vspace{-10pt}
        \begin{center}
            \subfloat[Rendered DOM code of element~\textcolor{box1color}{\textbf{\ding{182}}}.
            \label{fig:example_footer_after_dom}]{\includegraphics[width=1.0\linewidth, trim=225 201 225 201, clip]{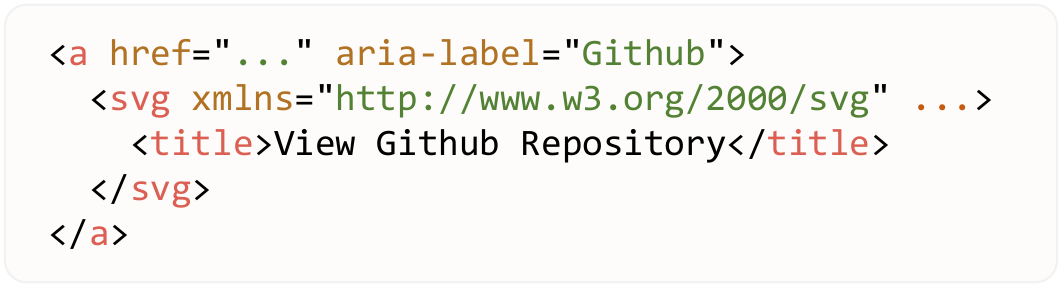}}
            \end{center}
    \vspace{-2pt}
        \begin{center}
            \subfloat[LLM's patch for Vio.~\textcolor{box4color}{\textbf{\ding{182}}} \textcolor{box4color}{\textbf{\ding{183}}} in Fig.~\ref{fig:example_footer}.
            \label{fig:example_footer_Conflict_Patch1}]{\includegraphics[width=1.0\linewidth, trim=258 146 258 96, clip]{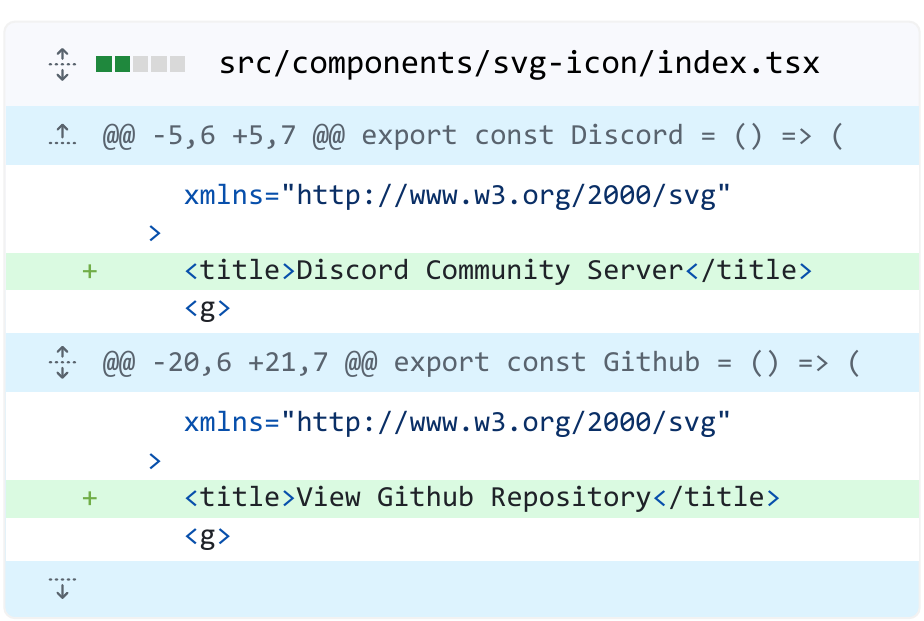}}
            \end{center}
        \begin{center}
            \subfloat[LLM's patch for Vio.~\textcolor{box4color}{\textbf{\ding{184}}} in Fig.~\ref{fig:example_footer}.
            \label{fig:example_footer_Conflict_Patch2}]{\includegraphics[width=1.0\linewidth, trim=258 183 258 160, clip]{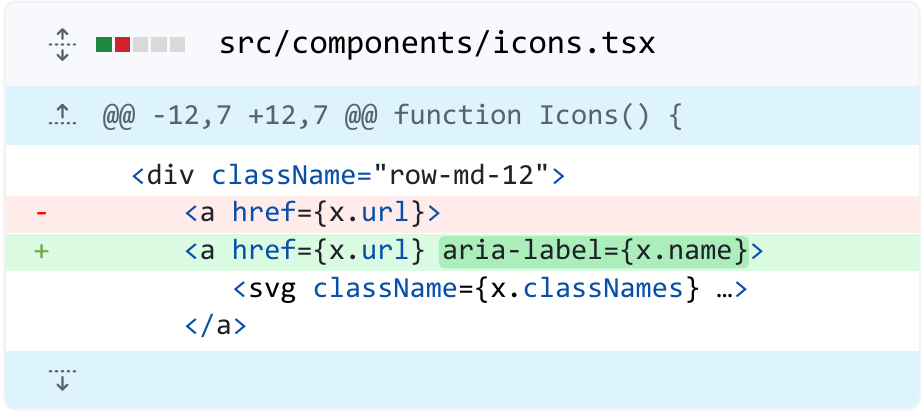}}
            \end{center}
    \vspace{-10pt}
    \caption{A motivating example of introduced side effects after fixing.} 
    \label{fig:knowledge_mining_example}
    \vspace{-10pt}
\end{wrapfigure}
\noindent
Figure~\ref{fig:example_footer_after} further shows that repairing the violations in Figure~\ref{fig:example_footer} individually introduces two violations
\textcolor{box1color}{\textbf{\ding{182}}} \textcolor{box1color}{\textbf{\ding{183}}}
caused by inconsistent names~\cite{Label_in_Name}.
As shown in Figure~\ref{fig:example_footer_after_dom}, this occurs when 
speech-input users navigate by speaking the visible text, a point of failure occurs when the visible label and the accessible name are different.
By analyzing the generated patches, we observe that the LLM first fixes two violations \textcolor{box4color}{\textbf{\ding{182}}} \textcolor{box4color}{\textbf{\ding{183}}} by adding \texttt{<title>} elements to individual SVG definitions (Figure~\ref{fig:example_footer_Conflict_Patch1}), but later resolves the third violation \textcolor{box4color}{\textbf{\ding{184}}} by adding an \texttt{aria-label} at the icon invocation level (Figure~\ref{fig:example_footer_Conflict_Patch2}).
This later global fix conflicts with earlier local fixes, producing unintended side effects.
Such conflicts arise because iterative repair lacks awareness of previously applied repair actions and component-level dependencies.

These observations suggest that individually repairing violations that share root causes or structural dependencies is often inefficient and error-prone in practice.
Effective web accessibility repair therefore requires grouping related violations and synthesizing coordinated repairs rather than treating each violation in isolation.

\begin{answerbox}
\small{
\textit{\textbf{Finding 2:} The Iterative Repair Strategy improves repair precision compared with the Basic strategy but substantially increases cost. Its unawareness of violation relationships leads to redundant operations or side effects, limiting efficiency and stability in complex accessibility tasks.}
}
\end{answerbox}

\subsubsection{WCAG Guided Repair Strategy}

This design aims to investigate whether incorporating WCAG knowledge via Retrieval Augmented Generation (RAG) improves repair.
As shown in Table~\ref{tab:wcag-strategy}, the WCAG-guided configuration slightly improves the resolve rate versus its non-guided counterpart (OpenHands$_{W}$ vs. OpenHands$_{W/O}$), but increases the overall computational overhead (\$64.86 to \$79.72).
A closer inspection reveals that WCAG guidance effectiveness depends on whether the violation requires standards-level reasoning.



\input{table/wcag_strategy_results_small.tex}
\noindent
For simple violations with explicit structural cues and well-known repair patterns (e.g., missing text alternatives), LLMs can synthesize fixes without external guidance.
For example, in the SVG icon case shown in Figure~\ref{fig:example_footer}, the model can directly add descriptive text to resolve the violation without consulting WCAG.
In such cases, blindly retrieving guideline excerpts only increases context length and cost, while potentially introducing noise that distracts the model from straightforward repairs.

\begin{wrapfigure}{l}{0.65\columnwidth}
    \vspace{-6px}
    \centering
    \includegraphics[width=1.0\linewidth, trim=0 0 0 0, clip]{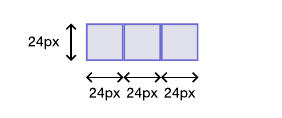}
    \vspace{-26px}
    \caption{The requirement is for targets to be at least 24 by 24 CSS pixels in size.}
    \Description{The requirement is for targets to be at least 24 by 24 CSS pixels in size.}
    \label{fig:target_size_basic}
    \vspace{-10px}
\end{wrapfigure}
\noindent
In contrast, complex violations require precise interpretation of WCAG \textit{Success Criteria}.
For instance, adjusting target size or spacing must satisfy \textit{SC~2.5.8 Target Size}~\cite{SC_2.5.8}, which explicitly requires pointer targets to be at least 24 by 24 CSS pixels.
As shown in Figure~\ref{fig:target_size_basic}, such repairs demand adherence to quantitative thresholds and implementation-specific constraints difficult to infer without domain knowledge.


These observations indicate that while WCAG knowledge is crucial for complex, standards-sensitive repairs, indiscriminately applying RAG to all violations is inefficient.
Traditional RAG strategies lack a mechanism for LLMs to assess whether domain knowledge is necessary for a given violation, leading to wasted context budget for simple cases and suboptimal knowledge utilization.

\begin{answerbox}
\small{
\textit{\textbf{Finding 3:}
WCAG is essential for repairing complex, standards-sensitive violations, but provides limited benefit for simple cases with deterministic fixes.
Blindly applying RAG to all violations increases cost and noise, highlighting the need for selective and context-aware knowledge invocation. 
}
}
\end{answerbox}

%% file: table/basic_and_iterative_result_small.tex
\begin{table}[t]
  \centering
  \vspace{6pt}
  \caption{Basic and Iterative Repair Strategy.}
  
  \label{tab:basic_and_iter_repair_strategy}
  \vspace{-4pt}
  \resizebox{\columnwidth}{!}{
    \begin{tabular}{lccr||lccr}
      \toprule
      \multicolumn{4}{c||}{\textbf{Basic Repair Strategy}} & \multicolumn{4}{c}{\textbf{Iterative Repair Strategy}} \\
      \multicolumn{1}{c}{\textbf{Systems}} & \textbf{$R_{\text{solve}}$} & \textbf{$N_{\text{side}}$} & \textbf{$C_{\text{total}}$} &
      \multicolumn{1}{c}{\textbf{Systems}} & \textbf{$R_{\text{solve}}$} & \textbf{$N_{\text{side}}$} & \textbf{$C_{\text{total}}$} \\
      \midrule
      SWE-agent$_{B}$   & 12.40\% & 0  & \$31.62  & SWE-agent$_{I}$   & 58.97\% & 13 & \$173.22 \\
      OpenHands$_{B}$   & 17.09\% & 1  & \$22.60  & OpenHands$_{I}$   & 61.53\% & 28 & \$64.86  \\
      GUIRepair$_{B}$   & 31.52\% & 5  & \$1.25   & GUIRepair$_{I}$   & 72.65\% & 41 & \$18.87  \\
      \bottomrule
    \end{tabular}
  }
\vspace{-6pt}
\end{table}

%% file: table/wcag_strategy_results_small.tex
\begin{wraptable}{r}{0.6\columnwidth}
\centering
\vspace{-12pt}
\caption{Knowledge Guided Strategy.}
\vspace{-10pt}
\label{tab:wcag-strategy}
\resizebox{0.6\columnwidth}{!}{
\begin{tabular}{lccr}
\toprule
\multicolumn{1}{c}{\textbf{Systems}} & \textbf{$R_{\text{solve}}$} & \textbf{$N_{\text{side}}$} & \multicolumn{1}{c}{\textbf{$C_{\text{total}}$}} \\ \midrule
OpenHands$_{W/O}$                     & 61.53\%                     & 28                         & \$64.86                                         \\
OpenHands$_{W}$                       & 69.23\%                     & 7                          & \$79.72                                        \\ \bottomrule
\end{tabular}
}
\vspace{-14pt}
\end{wraptable}

%% file: section/4_Approach.tex
\section{Approach}


In this section, we present A11YRepair, an LLM-based framework designed to efficiently repair web accessibility violations through a goal-oriented divide-and-conquer strategy integrated with domain knowledge from WCAG. Similar to most APR works, A11YRepair follows the basic repair workflow, including fault localization and patch generation. Unlike prior work, A11YRepair introduces a hierarchical planning mechanism that decomposes complex repair tasks into manageable subproblems and embeds accessibility-specific expertise to improve both repair effectiveness and efficiency.
As shown in Figure~\ref{fig:a11yrepair_workflow}, 
A11YRepair integrates five components into a unified end-to-end workflow. The \ding{182}-\ding{183}\textit{Violation Grouping Mechanism} structures the overall process by hierarchically organizing violations into component- and situation-level groups, enabling modular and scalable repair. The \ding{184}\textit{Fault Localization} and \ding{185}\textit{Patch Generation} modules execute the repair operations, identifying relevant bug files and generating targeted edits guided by both textual and visual cues. Throughout these stages, the \ding{186}\textit{WCAG-Driven Knowledge Integration} module dynamically determines whether domain expertise is needed and selectively retrieves relevant \textit{Success Criteria} and \textit{Techniques} to support reasoning on complex violations. Finally, all patches are applied to the webpage, and an accessibility checker re-evaluates the repaired version to measure improvement. 

\begin{figure}[t]
    \centering
    \vspace{3pt}
    \includegraphics[width=1.0\linewidth, trim=1058 640 1042 645, clip]
    {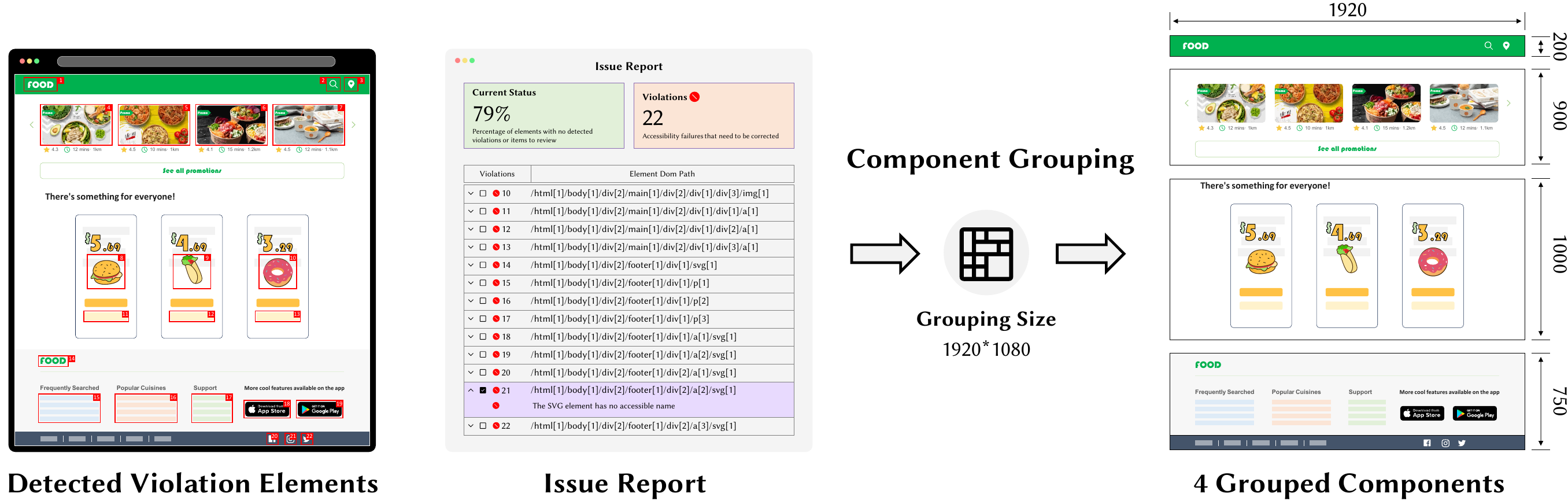}
    \vspace{-8pt}
    \caption{The area-based grouping strategy. 
    } 
    \Description{The size-based grouping strategy.}
    \label{fig:size_grouping}
    \vspace{-6pt}
\end{figure}

\subsection{Violation Grouping Mechanism }
\label{sec:Divide-and-Conquer Grouping Strategy}


As revealed in \textbf{Finding 1}, existing agent systems often lack effective planning capabilities for complex scenarios, making them not suited to handle large numbers of interdependent web A11Y violations.
In \textbf{Finding 2}, a naive divide-and-conquer strategy that feeds each violating element iteratively to the repair system can simplify the task but usually results in inefficient workflows and frequent side effects caused by redundant operations.
To address this limitation, A11YRepair adopts a goal-oriented divide-and-conquer strategy that compensates for the limited planning capacity of LLMs.
This strategy decomposes complex repair tasks into smaller, semantically coherent groups, allowing the system to balance repair cost and effectiveness more efficiently.
It operates in two coordinated phases: Fault Localization Grouping and Patch Generation Grouping.

\subsubsection{Fault Localization Grouping}
\label{sec:Fault-Localization-Grouping}

Inspired by the recent web generation work~\cite{DCGen}, A11YRepair employs a goal-oriented divide-and-conquer grouping strategy to organize A11Y violations into semantically coherent component-level groups for efficient fault localization, which includes two stages: a deterministic area-based initial grouping and an LLM-based refinement that adjusts grouping granularity based on the page’s structural and semantic layout.

\textbf{1) Area-based grouping.}
The area-based grouping stage provides a fast and language-agnostic method for deriving an initial set of component containers. As shown in Figure~\ref{fig:size_grouping}, after running the A11Y checker (e.g., IBM Accessibility Checker~\cite{IBM_A11YChecker}) on a webpage, A11YRepair collects all detected violations and their corresponding DOM paths. For each violation, the system traverses its DOM hierarchy from the root and selects the 
\kai{closest}
ancestor node whose rendered size falls below a predefined threshold (e.g., 1920×1080). All violations sharing the same ancestor are grouped together, while isolated elements without a suitable ancestor remain independent. 

\begin{figure}[t]
    \centering
    \vspace{3pt}
    \includegraphics[width=1.0\linewidth, trim=1050 640 1050 645, clip]
    {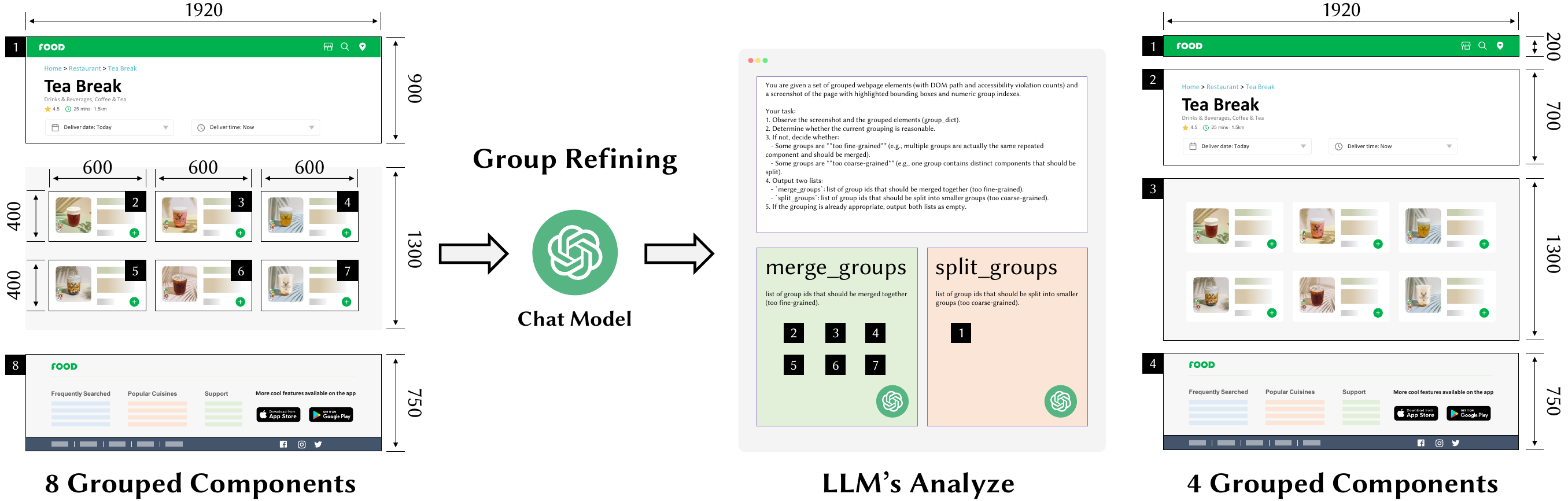}
    \vspace{-8pt}
    \caption{The llm-based refining strategy.} 
    \Description{The llm-based refining strategy.}
    \label{fig:llm_refining}
    \vspace{-6pt}
\end{figure}

\textbf{2) LLM-based refining.}
Rule-based grouping alone cannot generalize across the structural diversity of real-world websites. Some UI elements, such as compact navigation bars, are visually small yet semantically distinct from adjacent content and should not be merged with the main body. Conversely, repeated instances of the same component may appear as multiple small groups but should be treated collectively. To handle these cases, A11YRepair employs an LLM-based strategy to refine the initial grouping result.

As illustrated in Figure~\ref{fig:llm_refining}, the current grouping is rendered onto a webpage screenshot, where each group is visually highlighted and indexed with numeric identifiers. The model also receives structured metadata for every group, including violation types, DOM paths, and short DOM code snippets. The refinement objective is explicitly defined as grouping accessibility-violation elements that belong to the same UI component together. The LLM analyzes both the visual context and structural metadata to identify groups that should be merged (i.e., repeated or identical components) and those that should be split (i.e., overly broad clusters). 

Rather than performing direct merges or splits that could produce ambiguous or incorrect reassignments, A11YRepair adopts a conservative path-adjustment policy to implement the model’s recommendations. For groups to be merged, the grouping component path is adjusted upward by one DOM level (for example, from \texttt{/html[1]/body[1]/div[2]} to \texttt{/html[1]/body[1]}) so that broader ancestor nodes encompass multiple previously separate groups. For groups to be split, the grouping path is adjusted downward by one level to obtain finer granularity. This path-based adjustment effectively avoids combinatorial ambiguity and provides fine-grained control over grouping granularity.

As a single adjustment round may not yield the optimal structure, refinement is performed iteratively. After each adjustment, A11YRepair re-renders the updated grouping, presents the result to the LLM, and repeats the refine–adjust cycle until the model indicates no further changes or a preset iteration limit is reached. The final output is a set of refined component-level violation groups that capture both structural coherence and semantic distinctness. Each refined group is then treated as a single logical component and passed to the fault-localization module, where grouped violations are jointly analyzed to identify the relevant files for targeted repair.

By localizing violations at the component level rather than in isolation, this grouping strategy enables A11YRepair to reason about shared structure, styling, and behavior across related elements.
This design not only reduces redundant localization effort but also lays the foundation for root-cause–aware patch generation, where multiple related violations can be repaired coherently.

\subsubsection{Patch Generation Grouping}
\begin{figure}[h]
\vspace{-4pt}
    \centering
    \includegraphics[width=1.0\linewidth, trim=1122 655 1120 655, clip]
    {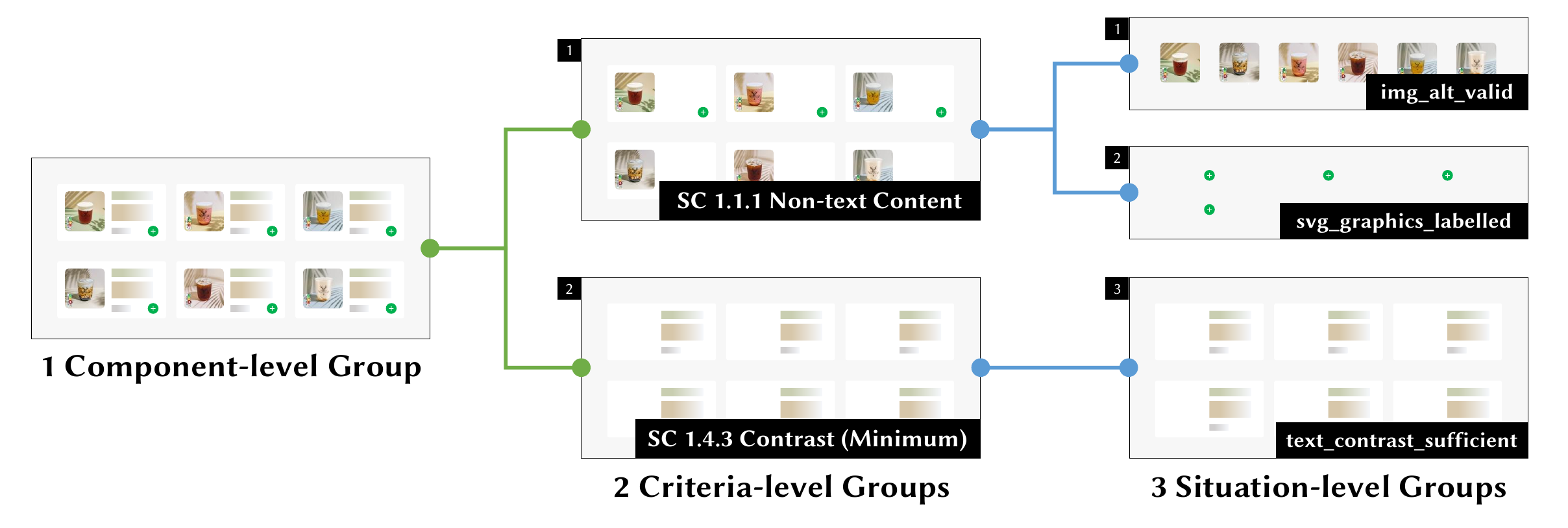}
    \vspace{-13pt}
    \caption{The 
    criterion/situation level 
    grouping strategy. 
    } 
    \Description{The patch generation (criterion/situation level) grouping strategy.}
    \label{fig:patch_gen_grouping}
\vspace{-6pt}
\end{figure}

After violations are localized into component-level groups, A11YRepair further refines these groups for patch synthesis.
While fault localization grouping determines \emph{where} to repair by identifying relevant components and files, it does not distinguish \emph{how} different violations within the same component should be repaired.
As highlighted by our motivating study (Finding~2), treating all violations within a component individually can lead to redundant edits and unintended side effects, especially when multiple violations stem from shared root causes.
Patch Generation Grouping addresses this issue by decomposing each component-level group into smaller, semantically coherent subsets, such that violations sharing the same underlying cause are repaired jointly.
As shown in Figure~\ref{fig:patch_gen_grouping}, the Patch Generation Grouping process proceeds in two hierarchical steps: 
criterion-based grouping and situation-based grouping.

\textbf{1) Criterion-based grouping.}
Given a component-level group obtained from the Fault Localization Grouping phase, all detected violations within the component are first clustered according to their violated WCAG \textit{Success Criteria} 
(i.e., Requirements~\cite{IBM_A11Y_Requirements}). 
This produces a set of criterion-level groups, each corresponding to a specific accessibility principle. For example, \textit{text\_contrast\_sufficient} violations map to \textit{SC~1.4.3: Contrast}, \textit{label\_name\_visible} violations correspond to \textit{SC~2.5.3: Label in Name}. These mappings derive from the A11Y checker’s rule–criterion associations (e.g., IBM Requirements~\cite{IBM_A11Y_Requirements_Rule}), allowing clustering of violations by criterion type.

\textbf{2) Situation-based grouping.}
Each criterion-level group is then refined into situation-level groups based on the specific contexts defined within WCAG SC. For instance, under \textit{SC~1.4.3 Contrast}~\cite{SC_1.4.3}, WCAG differentiates situations according to text size and weight:
\begin{answerbox_round}
\footnotesize{
\begin{itemize}[leftmargin=0.4cm]
\item \textit{Situation A:} Text smaller than 18pt (or 14pt if bold) must achieve a contrast ratio of at least 4.5:1 (\href{https://www.w3.org/WAI/WCAG22/Techniques/general/G18}{Technique~G18}).
\item \textit{Situation B:} Text of at least 18pt (or 14pt if bold) must achieve a contrast ratio of at least 3:1 (\href{https://www.w3.org/WAI/WCAG22/Techniques/general/G145}{Technique~G145}).
\end{itemize}
}
\end{answerbox_round}
\noindent
A11YRepair uses the violation messages returned by the accessibility checker to infer such situational contexts and perform message-level grouping automatically. By clustering violations into situation-level groups, the system decomposes complex, multi-intent repair tasks into smaller, semantically coherent subtasks. This design enables the LLM to generate focused, consistent, and WCAG-compliant patches guided by relevant WCAG Techniques~\cite{WCAG22_Techniques}. 

Importantly, this patch generation grouping strategy enforces \textit{root-cause–aware repair} by ensuring that violations sharing the same underlying cause are fixed jointly rather than individually.
By synthesizing a single coordinated patch for each situation-level group, A11YRepair avoids the conflicting or redundant edits that often arise when similar violations are repaired in isolation.

\subsection{Fault Localization and Patch Generation}
\label{sec:Fault Localization and Patch Generation}


After grouping all violating elements on the web page, we obtain Component-level and Situation-level Groups for fault localization and patch generation, respectively. Here, we draw inspiration from mainstream agentless repair systems' workflow design principles~\cite{Agentless,GUIRepair} to implement fault localization and patch generation. We also perform specialized design optimizations tailored to web bug repair scenarios to better map relationships between web elements and code files, while carefully avoiding conflicts introduced to the entire webpage by applying patch edits. Next, we will introduce the specific Fault Localization and Patch Generation process.

\subsubsection{Fault Localization}
The fault localization module leverages both the structural and visual characteristics of front-end web development to identify the source files responsible for accessibility violations. 
Unlike traditional back-end localization, which requires reasoning over control and data flows, front-end accessibility issues are tightly coupled with the visual presentation of UI components and their corresponding DOM identifiers (e.g., \texttt{id}, \texttt{class}, and text attributes). 
Existing APR systems, primarily designed for code-centric reasoning, often fail to utilize these front-end visual cues effectively~\cite{Agentless,ChatRepair,RepairAgent,FixAgent,SWE_agent,OpenHands}. 
To address this gap, A11YRepair introduces a reflective localization process that combines visual and textual evidence and iteratively refines its predictions, mimicking the hypothesis-testing process of human developers.

\begin{figure}[h]
    \centering
    \includegraphics[width=1.0\linewidth, trim=502 610 1272 610, clip]
    {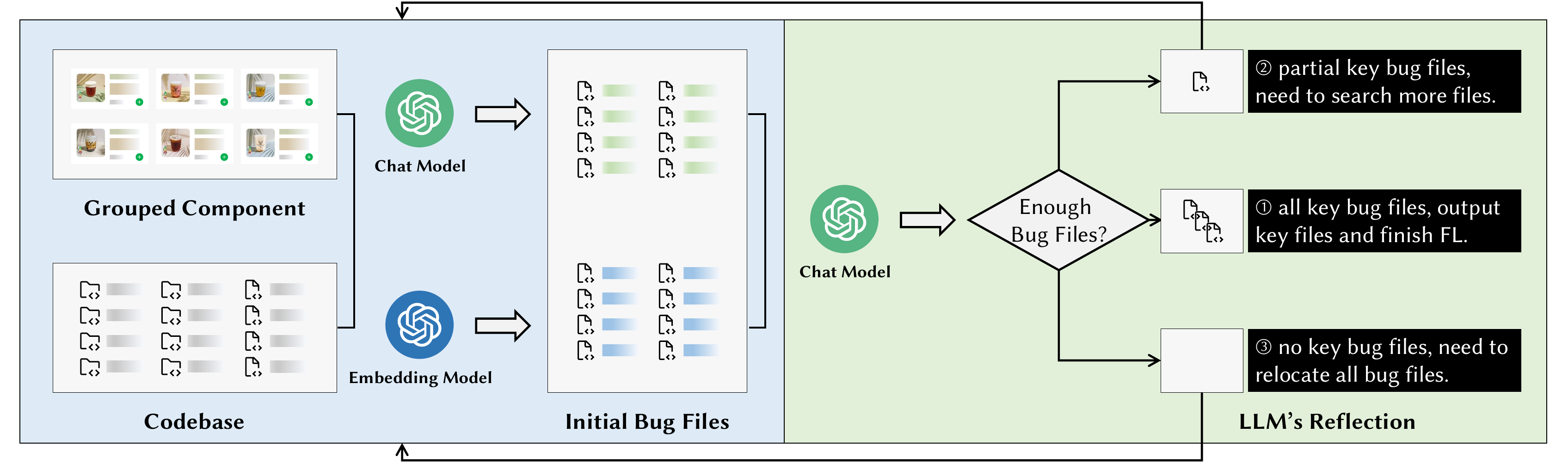}
    \vspace{-8pt} 
    \caption{The fault localization process, including File Localization (blue) and Locate Reflection (green).
    } 
    \Description{The fault localization process. \chen{Too small to see clearly...}}
    \label{fig:fault_localization}
    \vspace{-3pt}
\end{figure}

\textbf{1) File Localization.}
As shown in Figure~\ref{fig:fault_localization}, A11YRepair combines a chat model and an embedding model to identify potential bug files, following the general agentless paradigm~\cite{Agentless,GUIRepair}. 

\begin{itemize}[leftmargin=0.4cm]
    \item \textbf{Chat Model-based Localization:}
    The chat model receives structured violation-related information for a single component, including the A11Y bug report, relevant WCAG guidelines (if available), and annotated screenshots. 
    Following Agentless~\cite{Agentless}, we extract the repository structure and prompt the model to locate potential bug files by integrating its understanding of the violation data, visual cues, and repository layout. 
    To enhance contextual reasoning, we include both the violation types detected by the A11Y checker and the corresponding DOM tree code of each violated element. 
    A configurable slicing parameter $N$ determines how much surrounding DOM context to preserve. 
    For example, $N=1$ retains one level of ancestor and sibling nodes, while a larger $N$ extends the code context proportionally. 
    For visual input, we capture a predefined region (e.g., 1920×1080) of the target component; if it exceeds this size, the screenshot is dynamically resized to cover the full component area. 
    This strategy limits token consumption while maintaining full visibility of the relevant region. 
    Consistent with GUI testing practices, each violation element is indexed and labeled with a numeric identifier on the screenshot, allowing the model to directly associate frontend visuals with reported violations.

    \item \textbf{Embedding Model-based Localization:} 
    The embedding model complements the chat model by retrieving files using attribute-based similarity. 
    It extracts the DOM attributes of each violated element (e.g., \textit{ID}, \textit{CLASS}, \textit{TEXT}) and performs a symbol-matching search across the repository to identify the top-$N$ potentially related files. 
    Each file is ranked by the frequency of matching attributes, ensuring focus on UI-related code while filtering out unrelated files. 
    Because some configuration or build files may still be included, the chat model is again invoked to discard irrelevant entries before final retrieval. 
    The embedding model then computes semantic similarity between the issue report and the filtered candidate files, returning the top-$M$ most relevant ones. 
    This two-step process narrows the search space to files most closely aligned with the violated UI elements, substantially reducing retrieval costs.
\end{itemize}
Finally, A11YRepair merges the outputs of both models using a union operation to obtain the initial candidate set of bug files.

\textbf{2) Locate Reflection.}
To overcome the rigidity of single-pass localization in prior systems~\cite{Agentless,GUIRepair}, A11YRepair treats fault localization as a multi-turn reflective process. 
As shown in Figure~\ref{fig:fault_localization}, after producing an initial set of candidate bug files, the LLM iteratively evaluates whether these files contain the \textit{key bug files}—those that must be modified to resolve all accessibility violations. 
This reflection mechanism allows the model to analyze inter-file relationships and progressively refine its hypotheses, reducing the risk of missing critical files that contribute to the violation.

During each iteration, the model receives the violation-related information, WCAG guidelines, annotated screenshots, and initial candidate bug files. 
It is then prompted to reason about whether the set covers all, partial, or none of the required fixes. 
Based on this reasoning, the model executes one of three adaptive behaviors:
\begin{answerbox_round}
\footnotesize{
\begin{itemize}[leftmargin=0.4cm]
    \item \textbf{All Key Files Identified.}
    If the model concludes that the current candidate set already contains all key bug files, it outputs them as the final result and terminates localization.

    \item \textbf{Partial Coverage Detected.}
    If only part of the key files are found, the model restarts the localization phase using contextual cues from the current results to search for additional files.

    \item \textbf{No Relevant Files Found.}
    If none of the current candidates are relevant, the model revises its reasoning strategy and reinitializes the search from a new perspective.
\end{itemize}
}
\end{answerbox_round}
\noindent
This iterative reflection continues until the LLM determines that all key files have been identified or the max iteration limit is reached.

\subsubsection{Patch Generation}

With component-level violation groups localized to bug-related files, A11YRepair synthesizes concrete code edits to resolve issues. 
Since a component may contain multiple violations arising from different root causes, repairing them individually can introduce unintended side effects. 
To mitigate this risk, A11YRepair generates patches at the granularity of situation-level violation groups, jointly repairing violations with shared causes while handling distinct causes in separate rounds. 
This strategy reduces generation complexity and prevents fixes for one violation from interfering with others within the same component.

To ensure stable edits across potentially multiple files, A11YRepair emits patches in a structured \textit{search/replace} format~\cite{Agentless}, enabling deterministic modification locations. 
Each patch is applied incrementally and validated via a lightweight rebuild-and-render check, and is accepted only if the webpage renders correctly without introducing new console errors. 
This conservative validation loop avoids destructive changes and supports reliable patch applying.


\subsection{WCAG-Driven Knowledge Integration}
\label{sec:WCAG-Guided Knowledge Enhancement}

Web A11Y repair inherently depends on domain standards such as WCAG, which define \textit{Success Criteria} and \textit{Techniques} for detecting and resolving violations. 
However, as shown in \textbf{Finding 3}, simply injecting retrieved guidelines causes excessive context, redundant reasoning, and inconsistent knowledge usage. 
To address this, A11YRepair adopts a \textit{goal-oriented knowledge enhancement strategy} that selectively integrates WCAG guidance in a context-aware manner. 
Inspired by \textit{ReAct}~\cite{ReAct}, A11YRepair enables \textit{reflect before using knowledge}: it first analyzes whether the current situation warrants WCAG consultation, then decides what knowledge to apply. 
This reflection-before-knowledge process balances contextual relevance and computational efficiency.
A11YRepair implements this adaptive integration through a two-stage reflective decision process:

\begin{figure}[h]
  \centering
  {\includegraphics[width=0.95\linewidth, trim=960 393 960 393, clip]{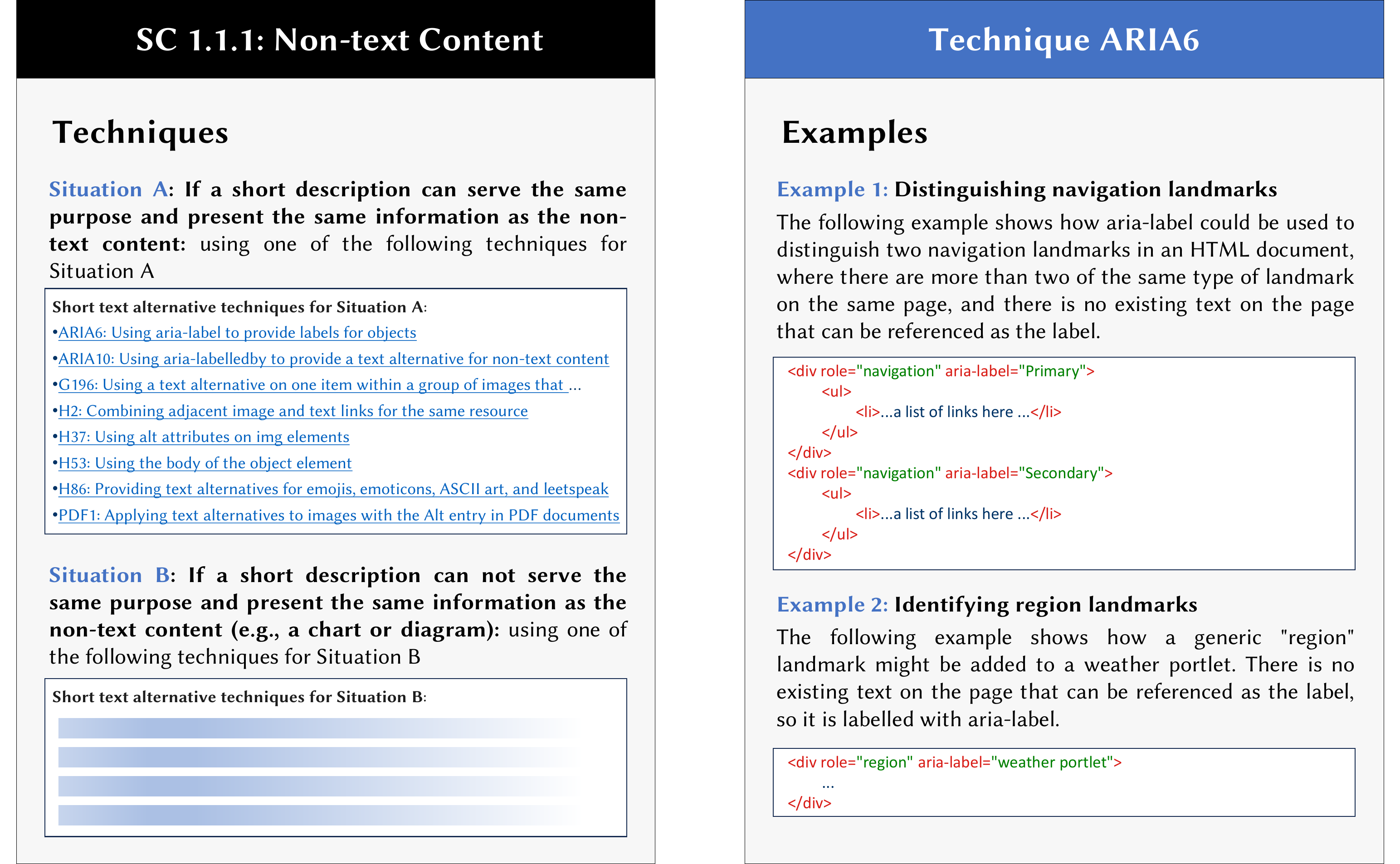}}
  \caption{Example of Success Criteria and Techniques.}
  \Description{Example of WCAG Success Criteria and Techniques.}
  \vspace{-6pt}
  \label{fig:sc_tec_example}
\end{figure}

\textbf{1) Necessity Analysis (Reflection on ``Whether'').}
Before incorporating WCAG knowledge, LLMs reflect on whether external guidance is needed for the current violation group. 
Given the A11Y bug report and annotated screenshot, the model assesses violation complexity. 
If the violations are simple (e.g., missing \texttt{alt} attributes), it outputs \texttt{"wcag\_required":"NO"} and skips enhancement. 
If the violations involve complex patterns (e.g., ARIA attributes, dynamic components, or keyboard interactions), it outputs \texttt{"YES"} and proceeds to next stage. 
This reflection step ensures that knowledge retrieval is invoked only 
when truly 
necessary.

\textbf{2) Technique Selection (Reflection on ``What'').}
Once WCAG knowledge is deemed necessary, A11YRepair retrieves specific \textit{Success Criteria} associated with violations (e.g., SC~1.1.1 \textit{Non-text Content}~\cite{SC_1.1.1} for missing alt text). 
The LLM is provided with the violation context and the relevant SC documentation, and it then identifies applicable techniques from the “Sufficient” and “Advisory” sections (e.g., \href{https://www.w3.org/WAI/WCAG22/Techniques/aria/ARIA6}{ARIA6}, \href{https://www.w3.org/WAI/WCAG22/Techniques/general/G73}{G73}, \href{https://www.w3.org/WAI/WCAG22/Techniques/html/H37}{H37}). 
Figure~\ref{fig:sc_tec_example} shows an example
under SC~1.1.1, 
where each technique specifies actionable repair strategies. 
The LLM outputs selected technique IDs along with explanatory reasoning. 
Finally, detailed SC and technique documentation are selected and merged into a curated knowledge context used to guide repair.

%% file: section/5_Setup.tex
\section{Experiment Setup}

\subsection{Research Questions}


\begin{answerbox_round}
\footnotesize{
\begin{itemize}[leftmargin=0.4cm]
    
\item 
\textbf{RQ1}: How effective is A11YRepair compared with other SOTA repair systems?  

\item 
\noindent
\textbf{RQ2}: How does each component contribute to the overall repair capability?  

\item 
\noindent
\textbf{RQ3}: Can A11YRepair generalize to unseen task instances and base models?  

\item 
\noindent
\textbf{RQ4}: How practical is A11YRepair in improving real-world web accessibility?
\end{itemize}
}
\end{answerbox_round}

\subsection{Benchmark and Implementation}

\begin{answerbox_round}
\footnotesize{
\begin{itemize}[leftmargin=0.4cm]
    
\item \textbf{Benchmark}: We evaluate all tools in A11YBench~\cite{A11YBench_link}, which is divided into two versions: 
\textit{Lite} contains 10 repositories, \textit{Full} contains all 60 repositories. 
We use the small-scale A11YBench-Lite to evaluate the overall effectiveness, and the larger-sample A11YBench-Full for conducting the generalizability study.

\item \textbf{Implementation}: We 
use \texttt{o4-mini-2025-04-16}~\cite{o4_mini} as the chat model and \texttt{text-embedding-3-small}~\cite{text_embedding_3_small} as the embedding model in the implementation of A11YRepair. More details 
and settings are available in our artifact~\cite{A11YRepair_link}.

\end{itemize}
}
\end{answerbox_round}

\subsection{Evaluation Metrics}
\label{sec:EvaluationMetrics}

\begin{answerbox_round}
\footnotesize{
\begin{itemize}[leftmargin=0.4cm]
    
\item \textbf{Violation Solve Rate}: 
Following prior work~\cite{Access,AccessGuru}, we measure repair effectiveness by comparing the number of violations detected before and after patch application. We use the IBM A11Y Checker to extract the violation sets $\mathcal{V}_{\text{before}}$ and $\mathcal{V}_{\text{after}}$. 
The solve rate is defined: $R_{\text{solve}} = (|\mathcal{V}_{\text{before}}| - |\mathcal{V}_{\text{after}}|) / |\mathcal{V}_{\text{before}}|$.

\item \textbf{Side Effect Count}: 
A good repair system should avoid introducing sides effects~\cite{APR_survey_huang,A11Y_IOS}.  
We therefore measure the number of newly created violations after repair,
a smaller $N_{\text{side}}$ reflects better stability and fewer unintended regressions:
$N_{\text{side}} = 
\left| \{ v \mid v \in \mathcal{V}_{\text{after}} \land v \notin \mathcal{V}_{\text{before}} \} \right|
\label{eq:side_effects}.$ 

\item \textbf{Token Usage Cost}: 
We do not measure runtime due to its dependence on external factors such as hardware and parallelism. To evaluate cost-efficiency, we compute the total LLM tokens consumed during end-to-end repair. 
Let $T_{\text{total}}$ denote total token usage and $P$ denote the per-token cost:
$C_{\text{total}} = T_{\text{total}} \times P
\label{eq:token_usage}.$
\end{itemize}
}
\end{answerbox_round}

%% file: section/6_Evaluation.tex
\section{Evaluation}

\subsection{RQ1: Overall Effectiveness}

We select GUIRepair~\cite{GUIRepair} as the primary baseline because it follows the same agentless pipeline ~\cite{Agentless} as A11YRepair and achieved the strongest performance in our motivating study and SWE-Bench leaderboard~\cite{Leaderboard}. 
Following that setup, we evaluate GUIRepair variants: \textit{Basic Repair Strategy}, \textit{Iterative Repair Strategy}, and \textit{WCAG Guided Repair Strategy}, denoted as GUIRepair$_B$, GUIRepair$_I$, and GUIRepair$_W$. 
Table~\ref{tab:main_result_lite} summarizes the overall results.
Overall, compared to best GUIRepair$_I$, A11YRepair improves the resolve rate by 4.96\%, reduces side effects by 44.52\%, and lowers token cost by 54.99\%. 
These results demonstrate the effectiveness of its divide-and-conquer design in balancing repair quality, stability, and cost.

\input{table/Main_Result_Lite}

\subsection{RQ2: Ablation Study}

\input{table/Ablation_Result_All}

In the ablation study, we aim to evaluate the contributions of A11YRepair’s design choices in enhancing the repair capability. 
Given that checking patches for syntax errors and performing visual inspections to filter invalid patches have been extensively applied in prior work, patch generation strategies are not explored as our primary design contribution. Therefore, 
We focus on evaluating how our key design choices of A11YRepair, including violation grouping, fault localization, and knowledge integration,
contribute to its performance. Due to the high cost of large-scale experiments, all ablation studies are conducted on A11YBench-Lite.

\textbf{1) Grouping Granularity.}
To examine the effect of divide-and-conquer granularity, we evaluate 3 variants:
\textbf{A11YRepair$_{com}$}, \textbf{A11YRepair$_{cri}$}, \textbf{A11YRepair$_{sit}$}. 
These variants adopt component / criterion / situation-level grouping, respectively, enabling us to analyze how different levels of decomposition influence the result.
Table~\ref{tab:ablation_result_all} 
shows a clear trend: finer-grained grouping consistently improves repair effectiveness while reducing side effects. 
Although finer-grained grouping incurs a modest cost increase, A11YRepair$_{sit}$ achieves the best overall trade-off.
These results indicate that finer-grained grouping yields more homogeneous and contextually coherent subproblems, reducing LLM reasoning ambiguity and alleviating side effects.

\begin{figure}[h]
\vspace{-6pt}
    \centering
    \begin{subfigure}{0.225\textwidth}  
        \centering
        \includegraphics[width=\linewidth, trim=1520 641 1528 643, clip]{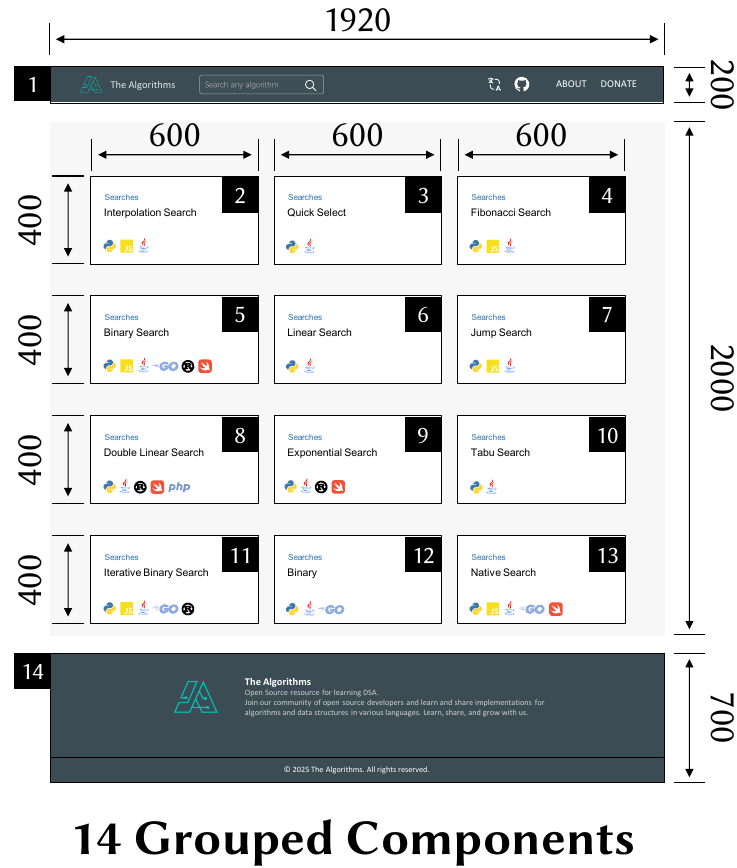}
        \vspace{-13pt}
        \caption{Grouping results before LLM-based refining.}
        \label{fig:refining_before}
    \end{subfigure}
    \hfill  
    \begin{subfigure}{0.225\textwidth}  
        \centering
        \includegraphics[width=\linewidth, trim=1520 641 1528 643, clip]{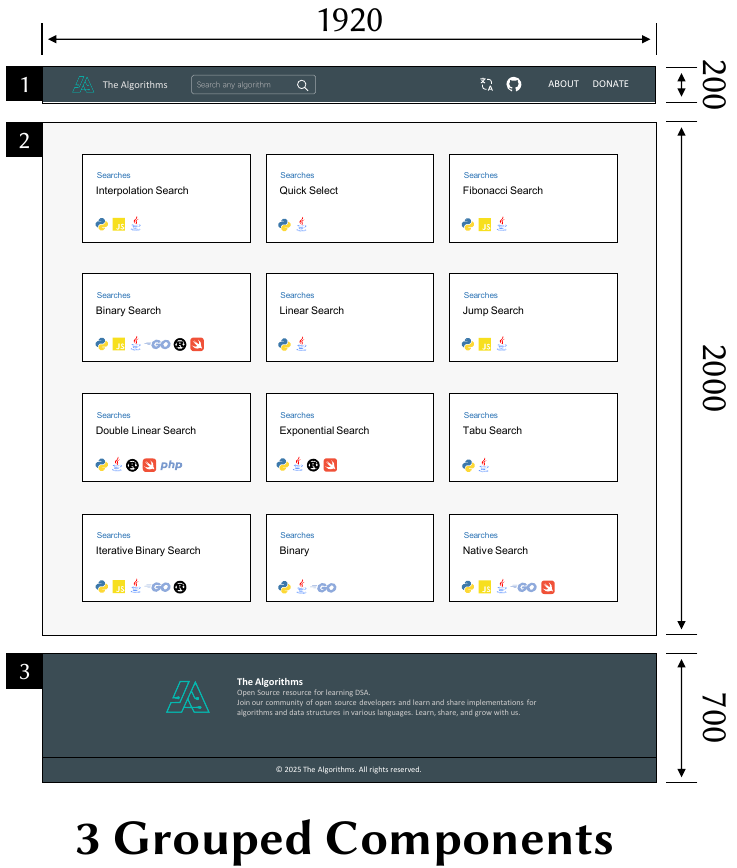}
        \vspace{-13pt}
        \caption{Grouping results after LLM-based refining.}
        \label{fig:refining_after}
    \end{subfigure}
    \vspace{-6pt}
    \caption{A case of using LLM-based refining strategy.}
    \label{fig:ablation_llm_refining}
    \vspace{-8pt}
\end{figure}

\textbf{2) LLM-based Refining.}
We further study the effect of LLM-based refining by comparing two variants: 
\textbf{A11YRepair$_{w/or}$} removes the refining step and uses only the area-based heuristic, whereas \textbf{A11YRepair$_{w/r}$} leverages the LLM to semantically adjust grouping decisions.
Table~\ref{tab:ablation_result_all}
shows that incorporating LLM-based refining yields substantial improvements. 
Figures~\ref{fig:ablation_llm_refining} presents a case from \href{https://github.com/TheAlgorithms/website.git}{\textit{Algorithms}}: the area-based grouping initially fragments several repeated algorithm-card components into separate groups (Figure~\ref{fig:refining_before}), whereas the LLM refining step correctly recognizes their shared semantics and merges them into a unified group (Figure~\ref{fig:refining_after}). Our further analysis shows that LLM refining merges 172 over-fragmented component groups and splits 6 over-clustered ones, eliminating redundant prompts and reducing unnecessary cost. By adaptively correcting heuristic grouping errors, the refining mechanism improves repair stability, prevents conflicting edits, and ensures consistent repairs across repeated components.

\begin{figure}[h]  
  \centering
  {\includegraphics[width=1.0\columnwidth, trim=1110 610 1110 590, clip]{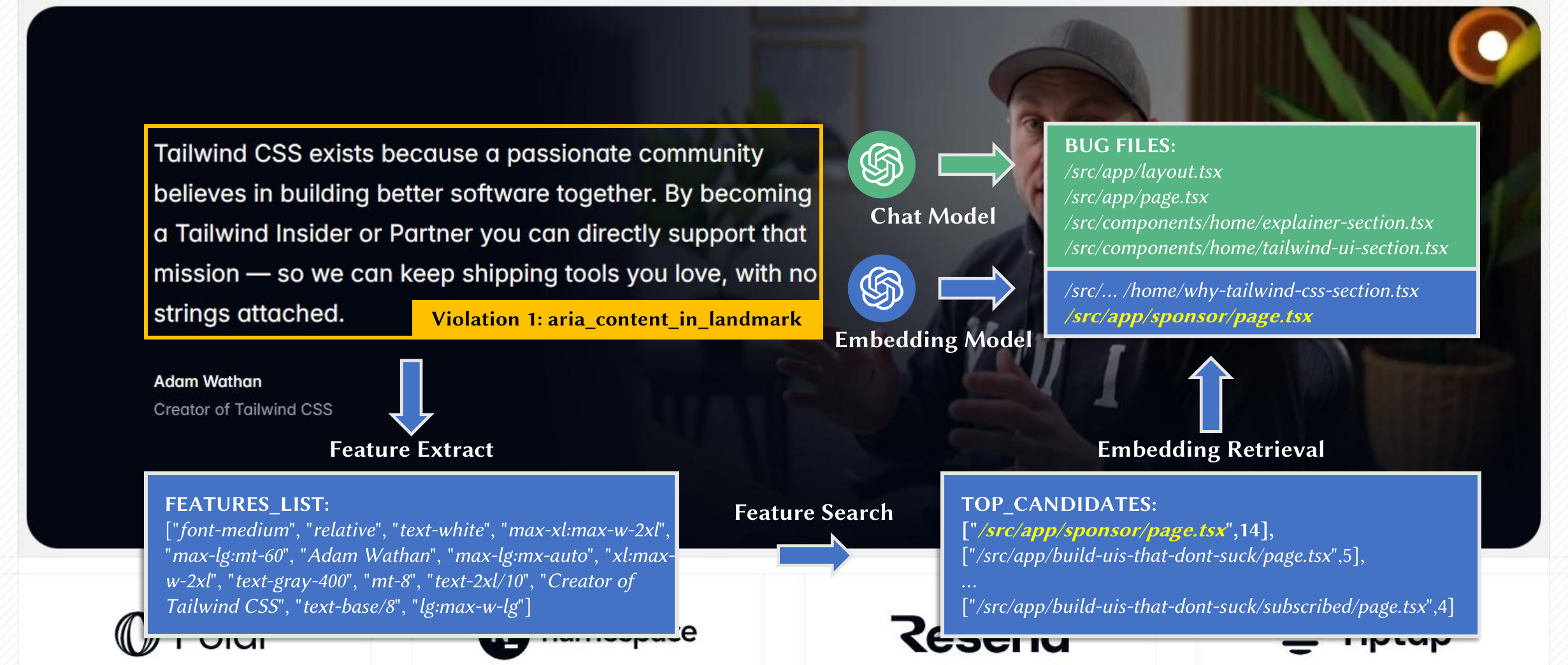}}
  \vspace{-14pt}
  \caption{A case of embedding model localization.}
  \Description{A case of embedding model localization.}
  \label{fig:embedding_case}
  \vspace{-4pt}
\end{figure}

\textbf{3) Feature Retrieval.}
During fault localization, A11YRepair extracts attribute-level cues from each violation element (e.g., component ID, key, and bounding box) to retrieve a focused set of structurally related candidate files. 
The embedding model then performs semantic retrieval within this reduced search space, alleviating the limitations of relying solely on a chat model, which often lacks structural grounding and may overlook files linked through UI feature correlations.
To assess its impact, we compare the full system with embedding-based retrieval (\textbf{A11YRepair$_{w/e}$}) against a variant that relies solely on the chat model (\textbf{A11YRepair$_{w/oe}$}).
Table~\ref{tab:ablation_result_all}
shows that incorporating feature retrieval (A11YRepair$_{w/e}$) yields a clear improvement over the chat-only variant (A11YRepair$_{w/oe}$).
To understand this improvement, we examine a case from \href{https://github.com/tailwindlabs/tailwindcss.com.git}{\textit{tailwindcss.com}} (Figure~\ref{fig:embedding_case}). 
The violation is a text block triggering the rule \textit{aria\_content\_in\_landmark}. Without feature retrieval, the chat model incorrectly predicts four unrelated files as potential bug files (\textit{layout.tsx}, \textit{page.tsx}, \textit{explainer-section.tsx}, \textit{tailwind-ui-section.tsx}). In contrast, the feature-guided embedding module correctly identifies \textit{/src/app/sponsor/page.tsx} as the true bug file by leveraging contextual attributes such as component IDs (FEATURE\_LIST).
These features allow the embedding model to rank the correct file first (TOP-1 in TOP\_CANDIDATES) without chat model inference. This shows that feature retrieval provides a fast, structurally grounded signal that complements the chat model and improves fault localization while reducing repair instability.

\textbf{4) Locate Reflection.}
To address incomplete initial localization, 
A11YRepair introduces a \textit{Locate Reflection} step that enables the LLM to verify whether the currently located files are sufficient.
We evaluate its effect by comparing a variant without the locate reflection (\textbf{A11YRepair$_{w/ol}$}) and the full version (\textbf{A11YRepair$_{w/l}$}).
As shown in 
Table~\ref{tab:ablation_result_all},
A11YRepair$_{w/l}$ improves the resolve rate by 11.9\% and reduces side effects by 53.89\% over A11YRepair$_{w/ol}$, with only minor token overhead. 
These gains arise because Locate Reflection both prunes irrelevant files from the initial localization and expands the candidate set when essential files are missing, transforming localization from a one-shot prediction into a self-corrective process.
\textbf{5) Knowledge Integration.}
We examined the impact of WCAG-driven knowledge integration strategy by 3 variants:
\textbf{A11YRepair$_{nw}$}, which removes all WCAG knowledge; 
\textbf{A11YRepair$_{aw}$}, which injects WCAG guidance for every instance; 
and \textbf{A11YRepair$_{sw}$}, which selectively incorporates WCAG content. 
This comparison allows us to assess whether external WCAG improves repair effectiveness over no knowledge at all, and whether selective, reflection-driven usage yields better results than unconditional, always-on knowledge injection.
As shown in 
Table~\ref{tab:ablation_result_all}, 
the selective-with-WCAG variant (A11YRepair$_{sw}$) delivers the best overall performance, improving the violation resolve rate by 3.21\% and reducing side-effect violations by 20 compared with the no-knowledge version (A11YRepair$_{nw}$), while also outperforming the always-with-WCAG variant (A11YRepair$_{aw}$) by a large margin in both resolve rate (78.15\% vs.\ 58.74\%) and side effects (172 vs.\ 498), with lower token cost as well (\$59.43 vs.\ \$75.79). These results reveal that WCAG knowledge is indeed helpful when applied selectively, since it provides targeted domain cues for complex violations that require explicit accessibility reasoning, yet unconditional WCAG injection often harms performance by introducing irrelevant or distracting context for simple violations the model can already solve.

\subsection{RQ3: Generalizability Study}




In this section, we evaluate whether our approach can be generalized to additional base models and unseen repositories.




\input{table/Gen_Result_Lite_small}
\textbf{1) Generalization across base models.}
To assess cross-model generalization, we further replace the base model 
while conducting experiments on A11YBench-Lite. 
As shown in Table~\ref{tab:gen_result_lite}, stronger models yield consistently better repair outcomes.
Using recent released models Gemini-3.0-flash and Kimi-K2.5, A11YRepair consistently maintains the resolution rate of around 80\%.
These results indicate that A11YRepair is model-agnostic and benefits from stronger base models: more capable reasoning engines amplify the effectiveness and cost-efficiency of the multi-stage repair pipeline.

\input{table/Main_Result_Full_small}
\textbf{2) Generalization across repositories.}
Table~\ref{tab:main_result_full} reports the results of A11YRepair on A11YBench-Full, it contains 60 repositories.
When using the {o4-mini}, A11YRepair achieves the highest resolve rate of 76.82\%, outperforming GUIRepair$_{I}$ by 16.68\% and GUIRepair$_{B}$ by 40.55\%. 
In terms of repair stability, A11YRepair introduces only 679 new side-effect violations, which is 71.51\% fewer than GUIRepair$_{I}$, and comparable to the lowest baseline GUIRepair$_{B}$. 
Furthermore, A11YRepair maintains a moderate repair cost of \$160.15, reducing cost by 66.63\% compared to GUIRepair$_{I}$ (\$160.15 vs. \$479.98). 
These results demonstrate that the divide-and-conquer repair strategy of A11YRepair remains robust and scalable, and that its advantages are even more pronounced on larger and more diverse project sets.



\subsection{RQ4: Usefulness Evaluation}
Evaluating only on benchmark is insufficient. We further investigate the usefulness of A11YRepair in real-world development settings: 



\input{table/Use_Result_Multi_Turns_small}
\textbf{1) Multi-Round Iterative Repair.}
In practice, A11Y repair is verification-driven and rarely one-shot: developers iteratively fix issues and rerun checkers to validate results.
To evaluate whether A11YRepair supports this workflow, we conduct a multi-round repair where checker feedback from each round guides subsequent repairs.
As shown in Table~\ref{tab:use_result_multi-turns}, A11YRepair steadily improves across rounds: after three iterations, the repair rate increases from 75.41\% to 87.49\%, while introduced side effects drop from 1.9\% to 0.6\%.
This shows that A11YRepair 
leverages post-repair verification to iteratively refine fixes, rather than relying on a single attempt.

\input{table/Manual_Check_small}
\textbf{2) Manual Patch Validation.}
Since checkers may produce false positives or overlook semantic issues, we manually audited the final repair results (Turn-3 in Table~\ref{tab:use_result_multi-turns}) to assess patch quality beyond checker signals.
Specifically, we inspected whether the reported violation was semantically resolved, and the patch preserved the intended UI functionality and layout. 
As shown in Table~\ref{tab:manual_check}, the manually validated resolve rate is comparable to, and slightly higher than, the checker-reported rate.
However, we also explicitly examined 6 silent regressions not captured by the checker: 
Five cases were related to contrast repairs that altered subtle UI color semantics (e.g., state-indicating differences under hover or dropdown interaction), which require dynamic interaction to detect. 
One case involved minor layout spacing changes following structural edits.
Although the observed silent regression ratio is low, these findings reveal two limitations of A11YRepair: 
\textit{dynamic UI state semantics} and
\textit{subtle layout-sensitive styling}.
Overall, manual audit confirms that A11YRepair produces semantically sound repairs under weak oracles, while also highlighting scenarios where human review remains beneficial.



\input{table/Use_Result_Github_PR_Top_50}

\textbf{3) Real-World Developer Feedback.}
To further assess the practicality of A11YRepair, we submitted patches generated by A11YRepair to GitHub projects. 
As a single web page may contain multiple 
violations and yield multiple
patch edits, we follow previous work~\cite{Iris} randomly selected 1 patch per project for submission to avoid bias.
Currently,
61 of submitted pull requests (PRs) have already been merged, 4 PRs have been approved. 
Due to space limitations, Table~\ref{tab:use_result_github_pr} shows only the top 50 merged PRs,
full lists please see~\cite{MergedPR_link}.
Notably, several patches were accepted by popular projects, including 
Google's \textit{\href{https://github.com/angular/angular}{angular}}, 
Microsoft's \textit{\href{https://github.com/microsoft/aspire.dev}{aspire}}, 
Facebook's \textit{\href{https://github.com/facebook/react-native-website}{react-native-website}}, 
IBM's \textit{\href{https://github.com/carbon-design-system/carbon-website}{carbon-website}}, 
Bytedance's \textit{\href{https://github.com/bytedance/deer-flow}{deer-flow}},
Ant Group's \textit{\href{https://github.com/ant-design/ant-design}{ant-design}}, 
and Alibaba's \textit{\href{https://github.com/spring-ai-alibaba/website}{spring-ai}} and \textit{\href{https://github.com/alibaba/hooks}{hooks}}. 
However, we also observed one rejection (\textit{\href{https://github.com/openclaw/openclaw.ai}{openclaw.ai} \#34}), where a contrast change was declined because the 
developer
preferred weaker visual contrast. This example is instructive: satisfying WCAG constraints does not automatically imply alignment with branding or aesthetic intent. Accessibility repair therefore benefits from human-in-the-loop collaboration, rather than fully autonomous deployment.
Overall, real-world adoption demonstrates that patches of A11YRepair are generally practical and developer-acceptable.

Furthermore,
we observed that A11YRepair can resolve violations rooted in third-party components.
For example, when repairing a violation reported in \textit{\href{https://github.com/ant-design/ant-design}{ant-design} \#56521}, A11YRepair identified that the violation element originated from an external component rather than the application itself.
Following developers’ guidance, we applied A11YRepair to the repository of the third-party component and generated a fix in \textit{\href{https://github.com/umijs/dumi}{dumi} \#2307}, which has been successfully merged.
This demonstrates that A11YRepair generalizes beyond single-application repair and can operate at framework level.



%% file: table/Main_Result_Lite.tex
\begin{table}[h]
\centering
\vspace{-5pt}
\caption{Repair Results on A11YBench-Lite (using o4-mini). 
}
\label{tab:main_result_lite}
\vspace{-5pt}
\resizebox{1.0\columnwidth}{!}{
\begin{tabular}{lc|rcr|rcr|rcr|rcr}
\toprule
  \multicolumn{1}{c}{\multirow{2}{*}{\textbf{Repo}}} &
  \multicolumn{1}{c}{\multirow{2}{*}{\textbf{Vio.}}} &
  \multicolumn{3}{|c|}{\textbf{GUIRepair$_{B}$}} &
  \multicolumn{3}{c|}{\textbf{GUIRepair$_{I}$}} &
  \multicolumn{3}{c|}{\textbf{GUIRepair$_{W}$}} &
  \multicolumn{3}{c}{\textbf{A11YRepair}} \\
  \multicolumn{1}{c}{} &
  \multicolumn{1}{c}{} &
  \multicolumn{1}{|c}{$R_{\text{solve}}$} &
  $N_{\text{side}}$ &
  \multicolumn{1}{c|}{$C_{\text{total}}$} &
  \multicolumn{1}{c}{$R_{\text{solve}}$} &
  $N_{\text{side}}$ &
  \multicolumn{1}{c|}{$C_{\text{total}}$} &
  \multicolumn{1}{c}{$R_{\text{solve}}$} &
  $N_{\text{side}}$ &
  \multicolumn{1}{c|}{$C_{\text{total}}$} &
  \multicolumn{1}{c}{$R_{\text{solve}}$} &
  $N_{\text{side}}$ &
  \multicolumn{1}{c}{$C_{\text{total}}$} \\ \midrule
\rowcolor[HTML]{EFEFEF}
\href{https://github.com/microsoft/TypeScript-Website.git}{TypeScript}              & 80   & 57.50\% & 0   & \$1.00 & 73.75\% & 14  & \$6.49   & \cellcolor[HTML]{807FFF}{\color[HTML]{FFFFFF}\textbf{85.00\%}} & 0   & \$6.40   & 81.25\% & 12  & \$2.54  \\
\href{https://github.com/golang/website.git}{golang}                  & 56   & 37.50\% & 0   & \$0.30 & \cellcolor[HTML]{807FFF}{\color[HTML]{FFFFFF}\textbf{69.64\%}} & 0   & \$3.60   & 57.14\% & 7   & \$3.99   & \cellcolor[HTML]{807FFF}{\color[HTML]{FFFFFF}\textbf{69.64\%}} & 0   & \$2.00  \\
\rowcolor[HTML]{EFEFEF}
\href{https://github.com/carbon-design-system/carbon-website.git}{carbon}          & 53   & 62.26\% & 2   & \$0.73 & 66.04\% & 0   & \$3.38   & 58.49\% & 14  & \$5.09   & \cellcolor[HTML]{807FFF}{\color[HTML]{FFFFFF}\textbf{84.91\%}} & 2   & \$1.26  \\
\href{https://github.com/tailwindlabs/tailwindcss.com.git}{tailwindcss}         & 225  & 59.82\% & 3   & \$0.41 & 70.09\% & 41  & \$9.15   & 40.00\% & 103 & \$18.82  & \cellcolor[HTML]{807FFF}{\color[HTML]{FFFFFF}\textbf{74.22\%}} & 50  & \$8.53  \\
\rowcolor[HTML]{EFEFEF}
\href{https://github.com/pnpm/pnpm.io.git}{pnpm.io}                 & 145  & 8.28\%  & 85  & \$0.60 & 66.90\% & 4   & \$6.63   & 68.97\% & 7   & \$11.29  & \cellcolor[HTML]{807FFF}{\color[HTML]{FFFFFF}\textbf{70.34\%}} & 0   & \$6.42  \\
\href{https://github.com/TheAlgorithms/website.git}{Algorithms}           & 634  & 29.81\% & 2   & \$0.53 & 79.18\% & 130 & \$22.03  & 86.75\% & 84  & \$35.07  & \cellcolor[HTML]{807FFF}{\color[HTML]{FFFFFF}\textbf{91.17\%}} & 27  & \$3.65  \\
\rowcolor[HTML]{EFEFEF}
\href{https://github.com/electron/website.git}{electron}                & 553  & 59.86\% & 86  & \$0.65 & 83.25\% & 23  & \$24.20  & 87.16\% & 14  & \$32.88  & \cellcolor[HTML]{807FFF}{\color[HTML]{FFFFFF}\textbf{89.33\%}} & 8   & \$8.72  \\
\href{https://github.com/lynx-family/lynx-website.git}{lynx}            & 504  & 22.82\% & 15  & \$0.69 & 60.98\% & 56  & \$21.21  & 51.59\% & 25  & \$19.40  & \cellcolor[HTML]{807FFF}{\color[HTML]{FFFFFF}\textbf{68.65\%}} & 43  & \$12.86 \\
\rowcolor[HTML]{EFEFEF}
\href{https://github.com/asyncapi/conference-website.git}{asyncapi}                & 79   & 1.98\%  & 1   & \$0.45 & 64.56\% & 22  & \$3.73   & \cellcolor[HTML]{807FFF}{\color[HTML]{FFFFFF}\textbf{82.28\%}} & 6   & \$1.75   & 65.82\% & 13  & \$1.47  \\
\href{https://github.com/PostHog/posthog.com.git}{posthog}             & 293  & 2.19\%  & 6   & \$0.89 & \cellcolor[HTML]{807FFF}{\color[HTML]{FFFFFF}\textbf{72.07\%}} & 20  & \$31.63  & 68.60\% & 31  & \$27.94  & 54.95\% & 17  & \$11.98 \\ \midrule
\rowcolor[HTML]{DFDCEF}
\multicolumn{1}{c}{\textbf{\#Total}} & 2622 & 32.22\% & 200 & \$6.24 & 73.19\% & 310 & \$132.03 & 71.66\% & 291 & \$162.63 & \cellcolor[HTML]{807FFF}{\color[HTML]{FFFFFF}\textbf{78.15\%}} & 172 & \$59.43  \\ \bottomrule
\end{tabular}
}
\vspace{-10pt}
\end{table}

%% file: table/Ablation_Result_All.tex
\begin{wraptable}{r}{0.55\columnwidth}
\vspace{-12pt}
\centering
\caption{Ablation study.}
\label{tab:ablation_result_all}
\vspace{-7pt}
\resizebox{0.55\columnwidth}{!}{
\begin{tabular}{lccc}
\toprule
\multicolumn{1}{c}{\textbf{Variants}} & \textbf{$R_{\text{solve}}$} & \textbf{$N_{\text{side}}$} & \textbf{$C_{\text{total}}$} \\
\midrule
\rowcolor[HTML]{807FFF}
\multicolumn{4}{c}{\color[HTML]{FFFFFF} {\ding{182}} Impact of Grouping Granularity}  \\
\rowcolor[HTML]{EFEFEF} A11YRepair$_{com}$    & 59.01\%  & 377  & \$43.75   \\
\rowcolor[HTML]{EFEFEF} A11YRepair$_{cri}$    & 64.91\%  & 288  & \$51.07   \\
\rowcolor[HTML]{EFEFEF} A11YRepair$_{sit}$    & 78.15\%  & 172  & \$59.43   \\
\midrule
\rowcolor[HTML]{807FFF}
\multicolumn{4}{c}{\color[HTML]{FFFFFF} {\ding{183}} Impact of LLM-based Refining}    \\
\rowcolor[HTML]{EFEFEF} A11YRepair$_{w/or}$   & 69.35\%  & 355  & \$108.76  \\
\rowcolor[HTML]{EFEFEF} A11YRepair$_{w/r}$    & 78.15\%  & 172  & \$59.43   \\
\midrule
\rowcolor[HTML]{807FFF}
\multicolumn{4}{c}{\color[HTML]{FFFFFF} {\ding{184}} Imapct of Feature Retrieval}     \\
\rowcolor[HTML]{EFEFEF} A11YRepair$_{w/oe}$   & 69.64\%  & 400  & \$52.06   \\
\rowcolor[HTML]{EFEFEF} A11YRepair$_{w/e}$    & 78.15\%  & 172  & \$59.43   \\
\midrule
\rowcolor[HTML]{807FFF}
\multicolumn{4}{c}{\color[HTML]{FFFFFF} {\ding{185}} Impact of Locate Reflection}     \\
\rowcolor[HTML]{EFEFEF} A11YRepair$_{w/ol}$   & 66.25\%  & 373  & \$46.16   \\
\rowcolor[HTML]{EFEFEF} A11YRepair$_{w/l}$    & 78.15\%  & 172  & \$59.43   \\
\midrule
\rowcolor[HTML]{807FFF}
\multicolumn{4}{c}{\color[HTML]{FFFFFF} {\ding{186}} Impact of Knowledge Integration} \\
\rowcolor[HTML]{EFEFEF} A11YRepair$_{nw}$     & 74.94\%  & 194  & \$53.84   \\
\rowcolor[HTML]{EFEFEF} A11YRepair$_{aw}$     & 58.74\%  & 498  & \$75.79   \\
\rowcolor[HTML]{EFEFEF} A11YRepair$_{sw}$     & 78.15\%  & 172  & \$59.43   \\
\bottomrule
\end{tabular}
}
\vspace{-6pt}
\end{wraptable}

%% file: table/Gen_Result_Lite_small.tex
\begin{wraptable}{r}{0.63\columnwidth}
\centering
\vspace{-12pt}
\caption{Repair results of A11YRepair with different base models.}
\vspace{-8pt}
\label{tab:gen_result_lite}
\resizebox{0.63\columnwidth}{!}{
\begin{tabular}{lccr}
\toprule
\multicolumn{1}{c}{\textbf{Base Models}} & \textbf{$R_{\text{solve}}$} & \textbf{$N_{\text{side}}$} & \multicolumn{1}{c}{\textbf{$C_{\text{total}}$}} \\ \midrule
GPT-4.1 mini     & 60.45\% & 461 & \$21.03  \\
o4-mini          & 78.15\% & 172 & \$59.43  \\
GPT-5 mini       & 77.93\% & 157 & \$21.03  \\
Gemini-2.5-flash & 75.41\% & 171 & \$34.16  \\
Gemini-2.5-pro   & 79.56\% & 101 & \$106.80 \\
Gemini-3.0-flash & 80.62\% & 125 & \$92.53  \\ 
Kimi-K2.5        & 80.63\% & 322 & \$28.80  \\ 
\bottomrule
\end{tabular}
}
\vspace{-10pt}
\end{wraptable}

%% file: table/Main_Result_Full_small.tex
\begin{wraptable}{r}{0.62\columnwidth}
\vspace{-8pt}
\centering
\caption{Repair results on A11YBench-Full with o4-mini.}
\label{tab:main_result_full}
\vspace{-8pt}
\resizebox{0.62\columnwidth}{!}{
\begin{tabular}{cccr}
\toprule
\multicolumn{1}{c}{\textbf{Repair Systems}} & \textbf{$R_{\text{solve}}$} & \textbf{$N_{\text{side}}$} & \multicolumn{1}{c}{\textbf{$C_{\text{total}}$}} \\ \midrule
GUIRepair$_{B}$ & 36.27\% & 456  & \$19.36  \\
GUIRepair$_{I}$ & 60.14\% & 2387 & \$479.98 \\
A11YRepair       & 76.82\% & 679  & \$160.15 \\ \bottomrule
\end{tabular}
}
\vspace{-4pt}
\end{wraptable}

%% file: table/Use_Result_Multi_Turns_small.tex
\begin{wraptable}{r}{0.61\columnwidth}
\centering
\vspace{-13pt}
\caption{Multi-round repair result (Gemini-2.5-flash). 
}
\vspace{-8px}
\label{tab:use_result_multi-turns}
\resizebox{0.61\columnwidth}{!}{
\begin{tabular}{cccc}
\toprule
\textbf{Repair Rounds} & \textbf{$R_{\text{solve}}$} & \textbf{$N_{\text{side}}$} & \multicolumn{1}{c}{\textbf{$C_{\text{total}}$}} \\ \midrule
Turn-1        & 75.41\% & 171          & \$34.16 \\
Turn-2        & 85.99\% & 85           & \$49.31 \\
Turn-3        & 87.49\% & 55           & \$59.85 \\
\bottomrule
\end{tabular}
}
\vspace{-12pt}
\end{wraptable}


%% file: table/Manual_Check_small.tex
\begin{wraptable}{r}{0.61\columnwidth}
\vspace{-12pt}
\centering
\caption{Repair results after manual checking.}
\label{tab:manual_check}
\vspace{-8pt}
\resizebox{0.61\columnwidth}{!}{
\begin{tabular}{ccccc}
\toprule
  \textbf{Evaluation}    &
  \multicolumn{1}{c}{\textbf{$V_{\text{before}}$}} &
  \multicolumn{1}{c}{\textbf{$V_{\text{after}}$}} &
  \multicolumn{1}{c}{\textbf{$R_{\text{solve}}$}} &
  \multicolumn{1}{c}{\textbf{$N_{\text{side}}$}} \\ \midrule
A11Y Checker    & 2622       & 328       & 87.49\% & 55           \\
Manual Check & 2433       & 277       & 88.61\% & 56          \\ \bottomrule
\end{tabular}
}
\vspace{-12pt}
\end{wraptable}

%% file: table/Use_Result_Github_PR_Top_50.tex
\begin{table}[]
\centering
\vspace{4pt}
\caption{Developer acceptance of A11YRepair's patches.}
\vspace{-2pt}
\label{tab:use_result_github_pr}
\resizebox{1.0\columnwidth}{!}{
\begin{tabular}{llrr||llrr}
\toprule
\multicolumn{1}{c}{\textbf{No.}} &
  \multicolumn{1}{c}{\textbf{Repo}} &
  \multicolumn{1}{c}{\textbf{Star}} &
  \multicolumn{1}{c||}{\textbf{PR ID}} &
  \multicolumn{1}{c}{\textbf{No.}} &
  \multicolumn{1}{c}{\textbf{Repo}} &
  \multicolumn{1}{c}{\textbf{Star}} &
  \multicolumn{1}{c}{\textbf{PR ID}} \\ \midrule
\rowcolor[HTML]{EFEFEF}
1  & \href{https://github.com/angular/angular}{angular/angular} & 100k  & \#66723 & 26 & \href{https://github.com/socketio/socket.io-website}{socketio/socket.io} & 343 & \#520  \\
2  & \href{https://github.com/ant-design/ant-design}{ant-design/ant-design} & 97.7k & \#56510 & 27 & \href{https://github.com/carbon-design-system/carbon-website}{carbondesign/web} & 319 & \#4784 \\
\rowcolor[HTML]{EFEFEF}
3  & \href{https://github.com/OpenHands/OpenHands}{OpenHands/GUI} & 69.3k & \#12728 & 28 & \href{https://github.com/pnpm/pnpm.io}{pnpm/pnpm.io} & 287 & \#743  \\
4  & \href{https://github.com/TryGhost/Ghost}{TryGhost/Ghost} & 52.1k & \#25975 & 29 & \href{https://github.com/OpenRefine/openrefine.org}{OpenRefine/org} & 154 & \#507  \\
\rowcolor[HTML]{EFEFEF}
5  & \href{https://github.com/bytedance/deer-flow}{bytedance/deer-flow} & 31.3k & \#844   & 30 & \href{https://github.com/kedacore/keda-docs}{kedacore/keda} & 149 & \#1700 \\
6  & \href{https://github.com/GitbookIO/gitbook}{GitbookIO/gitbook} & 28.7k & \#3934  & 31 & \href{https://github.com/lutris/website}{lutris/website} & 149 & \#750  \\
\rowcolor[HTML]{EFEFEF}
7  & \href{https://github.com/recharts/recharts}{recharts/recharts} & 26.8k & \#6872  & 32 & \href{https://github.com/apache/apisix-website}{apache/apisix} & 146 & \#1985 \\
8  & \href{https://github.com/umijs/qiankun}{umijs/qiankun} & 16.6k & \#3101  & 33 & \href{https://github.com/apache/doris-website}{apache/doris} & 126 & \#3327 \\
\rowcolor[HTML]{EFEFEF}
9  & \href{https://github.com/github/opensource.guide}{github/opensource} & 15.3k & \#3597 & 34 & \href{https://github.com/vergecurrency/vergecurrency.com}{verge/currency} & 110 & \#1265 \\
10 & \href{https://github.com/alibaba/hooks}{alibaba/hooks} & 14.9k & \#2892  & 35 & \href{https://github.com/apache/incubator-seata-website}{apache/seata} & 109 & \#1061 \\
\rowcolor[HTML]{EFEFEF}
11 & \href{https://github.com/ethereum/ethereum-org-website}{ethereum/ethereum} & 5.9k  & \#17139 & 36 & \href{https://github.com/microsoft/aspire.dev}{microsoft/aspire} & 109 & \#299  \\
12 & \href{https://github.com/kubernetes/website}{kubernetes/website} & 5.2k  & \#54097 & 37 & \href{https://github.com/vitessio/website}{vitessio/website} & 63  & \#2064 \\
\rowcolor[HTML]{EFEFEF}
13 & \href{https://github.com/docker/docs}{docker/docs} & 4.5k  & \#24013 & 38 & \href{https://github.com/kubeedge/website}{kubeedge/website} & 60  & \#758  \\
14 & \href{https://github.com/umijs/dumi}{umijs/dumi} & 3.8k  & \#2307  & 39 & \href{https://github.com/open-feature/openfeature.dev}{openfeature/dev} & 60  & \#1329 \\
\rowcolor[HTML]{EFEFEF}
15 & \href{https://github.com/vuejs/docs}{vuejs/docs} & 3.2k  & \#3327  & 40 & \href{https://github.com/spring-ai-alibaba/website}{spring-ai/website} & 60  & \#259  \\
16 & \href{https://github.com/facebook/react-native-website}{facebook/react-native} & 2.1k  & \#4956  & 41 & \href{https://github.com/asyncapi/conference-website}{asyncapi/website} & 47  & \#904  \\
\rowcolor[HTML]{EFEFEF}
17 & \href{https://github.com/rescript-lang/rescript-lang.org}{rescript-lang/rescript} & 1.9k  & \#1170  & 42 & \href{https://github.com/goharbor/website}{goharbor/website} & 45  & \#700  \\
18 & \href{https://github.com/ipld/ipld}{ipld/ipld} & 1.3k  & \#363   & 43 & \href{https://github.com/Milkdown/website}{milkdown/website} & 41  & \#250  \\
\rowcolor[HTML]{EFEFEF}
19 & \href{https://github.com/elementary/website}{elementary/website} & 1.3k  & \#3966  & 44 & \href{https://github.com/spiffe/spiffe.io}{spiffe/spiffe.io} & 31  & \#368  \\
20 & \href{https://github.com/php/web-php}{php/web-php} & 1.1k  & \#1787  & 45 & \href{https://github.com/remix-run/remix-website}{remix-run/remix} & 30  & \#373  \\
\rowcolor[HTML]{EFEFEF}
21 & 
\href{https://github.com/OverTheWireOrg/OverTheWire-website}{overthewire/website} & 1.1k  & \#173   & 46 & \href{https://github.com/rook/rook.github.io}{rook/github.io} & 25  & \#177  \\
22 & \href{https://github.com/istio/istio.io}{istio/istio.io} & 818   & \#17108 & 47 & \href{https://github.com/shorebirdtech/website}{shorebirdtech/web} & 19  & \#402  \\
\rowcolor[HTML]{EFEFEF}
23 & \href{https://github.com/mitre-attack/attack-website}{mitre-attack/attack} & 571   & \#564   & 48 & \href{https://github.com/google/trillian-website}{google/trillian} & 15  & \#129  \\
24 & \href{https://github.com/files-community/Website}{files-community/web} & 463   & \#826   & 49 & \href{https://github.com/mlflow/mlflow-website}{mlflow/mlflow} & 15  & \#429  \\
\rowcolor[HTML]{EFEFEF}
25 & \href{https://github.com/jhipster/jhipster.github.io}{jhipster/jhipster} & 345   & \#1550  & 50 & \href{https://github.com/SWE-bench/swe-bench.github.io}{SWE-bench/io} & 12  & \#44  \\ 
\bottomrule
\end{tabular}
}
\vspace{-10pt}
\end{table}

%% file: section/7_Threats.tex
\section{Threats to Validity}

\noindent
\textbf{Data Leakage.}
Unlike many APR benchmarks that derive tasks from GitHub pull request histories~\cite{Defects4J,SWEbench,SWEbench_M}, 
which risks exposing ground-truth fixes during LLM training.
A11YBench is constructed from violations detected by the A11Y Checker.
This design avoids leaking the ground truth fixes into model pretraining corpora. 

\noindent
\textbf{Weak Oracles.}
A11Y checkers
may introduce false positives, so provide weak repair oracles to threaten the validity.
We mitigate this risk by using IBM A11Y Checker~\cite{IBM_A11YChecker}, which offers good reported precision (97.5\%) and highest recall among rule-based checkers~\cite{GenA11y}. 
Importantly, the manual audit, low silent regression count, and real-world PR acceptance collectively provide independent validation signals beyond the single checker. The empirical evidence suggests that improvements are not merely checker-specific artifacts.

%% file: section/8_Related.tex
\section{Related Work}


\noindent
\textbf{Automated Program Repair.}
Traditional work including search-based~\cite{Genprog}, constraint-based~\cite{Semfix}, and template-based methods~\cite{TBar} have recently evolved toward LLM-driven systems~\cite{APR_survey_huang,APR_survey_yang,APR_survey_zhang}. 
For example, AlphaRepair~\cite{AlphaRepair}, ChatRepair~\cite{ChatRepair}, and Gamma~\cite{Gamma} demonstrate how pretrained knowledge can yield human-like fixes. 
Especially, recent efforts explore autonomous repair agents~\cite{RepairAgent,SWE_agent,Live_SWE_agent,OpenHands,AutoCodeRover,SpecRover,Agent4APR_empirical,TraeAgent,Adverintent,Prometheus}.
These systems mimic human debugging to implement the repair workflow.
Unlike the general purpose repair systems, A11YRepair is the first repo-level web A11Y repair tool.

\noindent
\textbf{Web Accessibility Detection.}
Web A11Y detection~\cite{Ma11y,Web_reflow_dect,Web_keyboard_a11y_detecting,A11YNAVIGATOR} 
follows standards such as WCAG~\cite{WCAG22}.
These guidelines enabled the development of checkers~\cite{IBM_A11YChecker,WAVE,Axe_Core,LightHouse}. 
While effective for syntactic or layout violations, static checkers struggle with semantic issues, e.g., whether an image’s alt text 
reflects its visual content.
Recent work~\cite{GenA11y,AccessGuru}
leverage LLMs to identify semantic issues. 
These advances 
provide a foundation
for our study: A11YRepair addresses the remaining challenge by 
turning
detected violations into actionable
fixes for the end-to-end accessible web development.

\noindent
\textbf{Web Accessibility Correction.} 
Early work relied on rule-based HTML/DOM transformations~\cite{A11Y_Corr_1,A11Y_Corr_2} or expert-in-the-loop methods \cite{A11Y_Corr_6,A11Y_Corr_7}, while others addressed specific violations using computer vision, such as alt-text generation~\cite{A11Y_Corr_4}.
However, these methods have limited coverage and robustness.
Recent studies leverage LLMs to assist web A11Y correction, including LLM-based DOM rewriting~\cite{Access,AccessGuru} and coding assistance for accessible design~\cite{CodeA11Y}.
Still, existing tools only work on rendered DOM/HTML, providing temporary patches that do not modify the underlying source code. 
\kai{It means that these tools cannot help developers address accessibility issues fundamentally. This also explains why our experiment did not adopt previous work.}
In contrast, A11YRepair performs \emph{source-level}, \emph{repository-wide} repair of A11Y violations. 
By combining divide-and-conquer grouping and selective WCAG-guided reasoning, A11YRepair delivers practical fixes in real-world projects.

\noindent
\textbf{Mobile Accessibility Enhancement.} 
Accessibility also have been studied in the mobile platform~\cite{TIMESTUMP,A11yScan,Android_A11Y_study,Motorease,Iris,AccessiText,COALA,gucharacterizing,Groundhog,OverSight,Mobile_A11y_Repair}.
Prior work mainly targets specific issue types, such as annotating UI elements~\cite{A11Y_annotation}, predicting semantic labels~\cite{A11Y_labels}, and repairing size~\cite{Mobile_A11y_Repair} or color~\cite{Iris} related accessibility issues.
Recent studies explore LLM for mobile A11Y.
Representative efforts include using multi-agent systems to generate repair suggestions for IOS development~\cite{A11Y_IOS}, and leverage visual reasoning to automate alt-text generation~\cite{A11Y_Alt}.
However, existing methods are largely platform-specific and problem-oriented, limiting generalizability.
Extending A11YRepair to mobile platforms is an key direction for future work.




%% file: section/9_Conclusion.tex
\section{Conclusion}
Web A11Y violations remain widespread in modern web systems.
By decomposing violations into goal-oriented subproblems and injecting domain-specific knowledge derived from WCAG, A11YRepair enables more reliable and scalable source-level repair.
We hope this work motivates further research on principled LLM-based program repair and contributes to building a more inclusive web ecosystem.


%% file: section/10_Data.tex
\section{Data Availability Statement}
Artifact is available via an online website~\cite{A11YRepair_link} and Figshare~\cite{Artifact_link}.